# Ion-Modulated Polyelectrolyte Complexation of DNA and Polyacrylic Acid from Molecular Dynamics Simulations

*Sisem Ektirici[1], Vagelis Harmandaris[1,2,3], Christos N. Likos[4], Terpsichori S. Alexiou[4]*

1. Computation-Based Science and Technology Research Center, The Cyprus Institute, Nicosia 2121, Cyprus
2. Department of Mathematics and Applied Mathematics, University of Crete, GR-71409 Heraklion, Greece
3. Institute of Applied and Computational Mathematics, Foundation for Research and Technology Hellas, IACM/FORTH, GR-71110 Heraklion, Greece
4. Faculty of Physics, University of Vienna, Boltzmanngasse 5, A-1090, Vienna, Austria

**ABSTRACT**

The formation of complexes between like-charged polyelectrolytes challenges conventional electrostatic intuition and highlights the central role of ions in mediating macromolecular organization. Here, we investigate the salt-dependent association of DNA with poly(acrylic acid) (PAA) using atomistic molecular dynamics simulations in NaCl, $MgCl_2$, and $CaCl_2$ solutions. A time-resolved state classification scheme, based on heavy-atom distance and hydrogen-bond formation, was applied to distinguish bound and unbound configurations, enabling quantitative analysis of how ion valency modulates complex stability and structure. The results reveal a clear hierarchy of association strength with $Ca^{2+}$ promoting persistent complex formation through direct inner-sphere coordination between DNA phosphates and PAA carboxylates, $Mg^{2+}$ mediating weaker, transient bridging interactions and $Na^{+}$ exhibiting only electrostatic screening action with negligible bridge formation. Structural analysis shows that multivalent ions not only enhance complex stability but also reshape the molecular organization of both macromolecules. $Ca^{2+}$ induces expansion of DNA and compaction of PAA within a strongly-bridged complex characterized by directional alignment and backbone-dominated binding, whereas $Mg^{2+}$ promotes more transient groove associations and $Na^{+}$ supports flexible, weakly correlated contacts. Our findings provide molecular-level insight into ion-specific mechanisms underlying polyelectrolyte organization and inform the design of responsive biomaterials and nucleic acid-based assemblies in multivalent ionic environments.

## 1. INTRODUCTION

Polyelectrolytes are macromolecules that bear a high density of ionizable groups along their backbone and dissociate into charged polymer chains when dissolved in polar solvents such as water[1]. Their charge is compensated by mobile counterions, giving rise to long-range electrostatic interactions that profoundly alter chain conformation, solubility, and dynamics compared with neutral polymers. Depending on the ionizable groups, polyelectrolytes can be classified as strong, with pH-independent dissociation, or weak, with charge controlled by solution pH and ionic strength[2]. These charged macromolecules are ubiquitous in nature and technology, spanning nucleic acids, glycosaminoglycans, and proteins, as well as synthetic species such as poly(acrylic acid) (PAA), poly(styrene sulfonate), and poly(ethyleneimine)[3,4]. Their behavior in solution is governed by a delicate balance of electrostatic repulsion, counterion condensation, chain entropy, counterion-release entropy, solvent medium, and chain flexibility, leading to striking responses such as coil–stretch transitions, complexation, and phase separation[1,5]. Because of this unique combination of polymer physics and electrostatics, polyelectrolytes have become central to diverse applications ranging from biomedical hydrogels[6–10] and drug delivery carriers to water treatment[11], coatings[12], and energy materials[13,14]. Beyond single-chain properties, polyelectrolytes exhibit rich multi-chain assembly phenomena, most notably the formation of polyelectrolyte complexes (PECs) driven by electrostatic attraction and counterion release. PECs are assemblies that form spontaneously when oppositely charged polyelectrolytes are mixed in water, a phenomenon first recognized in early studies of "complex coacervation" by Bungenberg de Jong et al., which developed into a modern framework over the past century[15,16]. Building on those observations, Voorn and Overbeek provided the first quantitative theory of complex coacervation, combining Flory–Huggins mixing with Debye–Hückel electrostatics to describe when polymer–rich liquid phases (coacervates) appear[17]. Thermodynamically, PEC formation is largely entropy-driven: association of oppositely charged chains releases their small counterions to the bulk, yielding a favorable counterion-release entropy that can dominate the free-energy balance even when enthalpy is modestly positive[18,19]. Depending on charge density, mixing ratio, ionic strength, pH, chain architecture, and salt type, the products span a range from solid complexes to liquid coacervates to fully dissolved solutions[16,20].

At the molecular level, the formation of oppositely charged PECs is governed primarily by electrostatic attraction and the concomitant release of condensed counterions, which provides a substantial entropy gain that drives complexation even when the direct Coulomb interaction is partly screened by the surrounding counterions. Secondary contributions such as hydrogen bonds, hydrophobic contacts, and dipole interactions can further stabilize the assemblies, particularly when one or both polymers are weak polyelectrolytes or bear amphiphilic side chains[4]. Structural models for the resulting complex range from the ladder-like arrangement[21] of nearly stoichiometric complexes with ordered ion-pairing between chains to the more disordered "scrambled-egg" model[22], in which chains interpenetrate irregularly with residual bound counterions. The balance between electrostatic correlation, counterion entropy, and salt screening thus determines whether mixing a polycation with a polyanion produces a molecular complex, a colloidal dispersion, or a dense coacervate phase[23].

A natural extension of this classical picture considers what happens when the interacting polymers bear the same charge. Although mean-field electrostatics predicts mostly repulsive effective interactions between like-charged macromolecules[24,25], a substantial body of experimental and theoretical work has revealed conditions under which this repulsion can be overcome. Early studies on DNA condensation showed that multivalent cations such as spermidine and cobalt(III) hexamine can drive the attraction and collapse of highly charged DNA helices, demonstrating the role of counterion correlations and release in mediating like-charge association[26–28]. Investigations on filamentous actin further confirmed that divalent and trivalent cations induce bundle formation by reducing electrostatic repulsion and generating correlated ion-mediated forces between the negatively charged filaments[29]. More recently, studies of protein–polyelectrolyte systems have shown that heterogeneous surface charge distributions and local patches of opposite sign can promote binding even when the overall net charges are the same[30], and experiments inspired by mussel adhesive proteins revealed that strong cation–$\pi$ interactions can enable coacervation of highly cationic polymers despite their like charges [31]. Together, these findings demonstrate that mechanisms beyond mean-field electrostatics, including counterion correlations, ion-release entropy, charge patchiness, and specific short-range interactions, can generate effective attractions strong enough to reverse the repulsion predicted for like-charged macromolecules.

Recent work has shown that salt identity and ion valency strongly influence the behavior of charged macromolecular complexes. For oppositely charged synthetic polyelectrolytes, Perry et al. demonstrated that increasing monovalent salt concentration reduces coacervate formation in polyacrylate–polyallylamine systems, with the extent depending on the specific salt added[32]. Iyer et al. found that divalent salts such as $MgCl_2$ and $CaCl_2$ alter both the composition and the viscoelastic properties of polyelectrolyte complexes. Multivalent ions can also drive reentrant condensation of polyelectrolytes, as shown by Yong et al. who reported collapse and redissolution transitions arising from strong multivalent ion adsorption[33]. Ion specific effects among divalent ions appear in multiple systems. Braide et al. showed that $Ca^{2+}$ and $Mg^{2+}$ produce different glass transition temperatures in polyelectrolyte complexes, indicating a distinct internal organization[34]. In DNA systems, the ion type and the ionic strength have been shown to modulate polymer conformations and interchain interactions; Alexiou et al. used atomic simulations to show that monovalent and divalent counterions produce distinct conformational responses in DNA oligo-catanes with ionic strength dependent elongation and contraction regimes[35]. Single stranded DNA also displays ion specific behavior, with Liu et al. reporting that $Ca^{2+}$ leads to irreversible phase separation of DNA droplets, while $Mg^{2+}$ produces reversible transitions[36]. A recent study on multivalent ion–mediated polyelectrolyte association reported that $Ca^{2+}$ promotes PAA–PAA attraction through the formation of PAA–$Ca^{2+}$–PAA bridges and enhanced $Ca^{2+}$ coordination to the polymer, demonstrating a clear ion-specific mechanism for like-charge association[37].

Although substantial progress has been made in elucidating how ions regulate biomolecular and polymer complexation, existing studies have largely focused either on oppositely charged polyelectrolytes or on DNA interacting with specific multivalent cations. In contrast, heterogeneous assemblies composed of chemically dissimilar, like-charged polymers have received considerably less attention. In particular, the association of a highly charged, semi rigid biopolymer such as DNA with a flexible anionic polymer such as PAA has not been systematically examined.

Poly(acrylic acid) (PAA) exhibits a combination of properties, including high negative charge density, low toxicity, biocompatibility, and biodegradability, that make it an attractive component of a wide range of nanotechnological platforms. Emerging applications of PAA include its use as a functional coating in nanoparticle-based systems for DNA isolation and purification, as a building block in biosensing platforms, and as a polyelectrolyte layer enabling controlled DNA encapsulation and release[38–41]. DNA provides a rigid, anisotropic charge lattice with well-characterized binding sites[42], while PAA contributes a flexible, mobile chain whose interactions depend strongly on local ion coordination[43]. Complexes between DNA and synthetic polyanions are therefore relevant both as model systems for understanding like-charged macromolecular association and as functional assemblies in technologies in which nucleic acids are processed, transported, or immobilized in charged polymer environments. In such contexts, ion-mediated attraction between similarly charged species can determine the stability, selectivity, and responsiveness of the resulting materials.

The influence of counterion valency and size on bridging between dissimilar anionic chains is not well understood, and the atomistic organization of ions at DNA–polymer interfaces remains poorly characterized. To address this gap, the present work investigates DNA–PAA interactions in aqueous solution using atomistic molecular dynamics (MD) simulations. We compare systems containing monovalent sodium chloride (NaCl), small divalent magnesium chloride ($MgCl_2$) and larger divalent calcium chloride ($CaCl_2$) in order to resolve how ion valency and ionic size influence effective attraction and bridging formation between the two negatively charged polymers. These ions were chosen because they differ strongly in valency, ionic radius, and hydration dynamics, enabling systematic assessment of how these physicochemical properties regulate ion-mediated attraction and bridging between like-charged polymers.

This study provides a detailed molecular picture of similar charge complexation between DNA and PAA, clarifying how counterion-mediated correlations and ion size effects can promote attraction despite net negative charge. The results establish atomistic descriptors for association and structural response that can guide both theoretical models and future experiments on anion–anion complexation in biological and synthetic systems. We primarily evaluate the bound fraction ($f_{bound}$) for DNA–PAA association, defined as the time-averaged fraction of PAA units in contact with DNA within a standard distance cutoff. In addition, we assess the conformational response of DNA, PAA, and the complex they form by analyzing their structural parameters in each ionic environment. By providing an atomistic view of how counterion valency and size modulate the association of two anionic polymers, our work extends polyelectrolyte complexation beyond the conventional focus on oppositely charged polymers that form complexes through direct electrostatic attraction. These insights can inform the design of ion-responsive nucleic acid–polymer assemblies, guide experimental strategies for controlling anion–anion interactions in soft materials, and support the development of polyelectrolyte systems for applications such as gene delivery, biosensing, and advanced water treatment.

The following sections describe the simulation methodology and analysis framework used to elucidate salt-dependent DNA–PAA complexation. The Methodology section details the system construction, force-field parameters, simulation protocol, and analysis procedures employed to characterize binding, structure, and ion coordination. The Results and Discussion section then presents the salt-dependent association behavior, including bound-state classification, interaction energies, ion distributions, bridging statistics, contact patterns, and orientational organization of the complexes. The final section summarizes the main findings and discusses their implications for polyelectrolyte theory and ion-responsive macromolecular assembly.

## 2. METHODOLOGY

The molecular mechanisms underlying the interaction between DNA and PAA in different ionic environments were explored through atomistic MD simulations of DNA in the absence and presence of PAA. This study was designed to (i) characterize how monovalent and divalent salts modulate DNA–PAA complex formation and (ii) quantify the structural response of DNA, PAA, and the resulting complex under different ionic environments. Three salt types were considered: NaCl, $MgCl_2$, and $CaCl_2$. DNA-only and PAA-only systems served as references to isolate the structural effect of the polymer.

## a. System Construction and Simulation Protocol

All DNA systems contained a single 24 base pair DNA duplex with one strand having the sequence 5′–ATCG–3′, repeated six times and complementary strand 5′ -TAGC-3′ repeated six times, representing a linear B-DNA fragment. This artificial sequence was chosen as a simple, compositionally balanced model fragment, with equal numbers of A T and C G base pairs and no specific biological motif. The goal was to use a generic double stranded DNA segment of moderate length that is long enough to exhibit typical groove geometry but still feasible for extensive atomistic simulations. The full list of base pairs used in the duplex is provided in the Figure S1. The fully atomistic duplex comprises 1522 atoms and carries a net charge of $-46\ e^{-}$. DNA–PAA systems contained the same duplex together with one PAA chain of 40 monomer units. The PAA chains were modeled with 50% of the carboxyl groups deprotonated, with the deprotonation pattern assigned randomly along the chain, corresponding to a pKa of 4.5[44] and resulting in a net charge of $-20\ e^{-}$. Atomic charges were defined by the employed force field, resulting in nonuniform charge distributions along both the DNA and PAA molecules. The AMBER parmBSC1 force field was used for DNA which incorporates refined $\varepsilon$, $\zeta$ backbone dihedral parameters along with the $\chi$ glycosidic torsion parameters[45]. ParmBSC1 has been extensively validated and shown to reproduce DNA conformational properties and stability with improved accuracy compared to earlier AMBER force fields[46–48]. The PAA chains were parametrized using the Generalized Amber Force Field (GAFF), which is commonly applied to synthetic polymers[49–51]. Each system was solvated in a cubic box of 14.8 nm per side with explicit TIP3P water molecules[52]. A physiological ionic strength of I = 0.15 M was obtained by adding the corresponding numbers of free ions ($Na^{+}$ and $Cl^{-}$ in NaCl systems, $Mg^{2+}$ and $Cl^{-}$ in $MgCl_2$ systems, and $Ca^{2+}$ and $Cl^-$ in $CaCl_2$ systems). Ions were added individually rather than as neutral salt molecules. Additional counterions were introduced as needed to neutralize the net charge of DNA and/or PAA. Parameters for $Na^+$, $Mg^{2+}$, and $Ca^{2+}$ ions were taken from the Joung and Cheatham parameter set, which is optimized for the TIP3P water model to reproduce experimental ion–water interactions and solvation thermodynamics[53,54]. DNA-only systems were constructed under the same conditions to serve as references for evaluating polymer-induced structural changes. Table 1 summarizes the composition and size of the simulated systems.

**Table 1.** Number of atoms for simulated systems for DNA, PAA and DNA–PAA complexes under different ionic conditions. (# stands for Total Number of)

| System | #Atoms | # Water | # $Ca^{+2}$/$Mg^{+2}$ | # $Na^+$ | # $Cl^-$ |
|---|---|---|---|---|---|
| $[DNA]_{NaCl}$ | 314649 | 312495 | - | 339 | 293 |
| $[DNA]_{CaCl2}$ | 314109 | 311685 | 316 | - | 586 |
| $[DNA]_{MgCl2}$ | 314133 | 311709 | 316 | - | 586 |
| $[PAA]_{NaCl}$ | 46309 | 15287 | - | 63 | 43 |
| $[PAA]_{CaCl2}$ | 46243 | 15254 | 53 | - | 86 |
| $[PAA]_{MgCl2}$ | 46243 | 15254 | 53 | - | 86 |
| $[DNA\text{-}PAA]_{NaCl}$ | 314056 | 311280 | 326 | - | 586 |
| $[DNA\text{-}PAA]_{CaCl2}$ | 314576 | 312060 | - | 359 | 293 |
| $[DNA\text{-}PAA]_{MgCl2}$ | 314056 | 311280 | 326 | - | 586 |

All systems were first energy-minimized using the steepest descent algorithm to remove steric clashes and relax unfavorable contacts. This was followed by a two-stage equilibration protocol: an NVT equilibration for 200 ps to stabilize temperature, followed by a 200 ns NPT equilibration to reach the target density at 1 bar. Temperature was maintained at 300 K using the velocity-rescale thermostat with separate coupling groups for solute and solvent, to avoid thermostat-induced distortion of macromolecular dynamics arising from the much faster relaxation of water and ions. Separate coupling ensures that DNA and PAA experience physically meaningful thermal fluctuations while the solvent remains properly equilibrated. Pressure was controlled with the stochastic cell-rescale barostat using isotropic coupling. Electrostatics were treated using the Particle-Mesh Ewald (PME) method[55] with a real-space cutoff of 1.4 nm, with direct Coulomb interactions computed only up to this distance and the remaining long-range terms evaluated in reciprocal space. The PME grid spacing was 0.12 nm with fourth-order interpolation. Van der Waals interactions were truncated at 1.0 nm, and periodic boundary conditions were applied in all three directions.

Production simulations were carried out in the NPT ensemble using the leapfrog integrator with a 1 fs time step. DNA-only systems were simulated for 500 ns, while DNA–PAA systems were simulated for 1 μs. DNA and PAA were inserted into the solvated box in non-interacting configurations, with a minimum initial separation of about 4 nm to prevent any artificial bias toward complex formation. To improve statistical reliability, three independent replicas were performed for each DNA–PAA system, starting from distinct initial velocity distributions.

### b. Definition and Quantification of DNA-PAA Complexation

The interaction between DNA and PAA differs fundamentally from typical biomolecule–polymer systems because both partners are highly charged polyelectrolytes**.** Unlike the binding of many biomolecules to synthetic polymers, which is often mediated by localized charged or hydrophobic surface patches, any potential interaction between DNA and PAA arises from the interplay of long-range electrostatic attraction and repulsion, counterion condensation, and dynamic screening by the salt environment[56,57]. Contacts in such systems are highly dynamic, chains may associate, partially unwrap, and rebind repeatedly as local charge fluctuations and ion rearrangements occur. Consequently, a single, well-defined equilibrium complex is less meaningful, and transient encounters cannot be interpreted as stable binding events. To obtain a quantitative and physically meaningful definition of complexation, we applied a persistence-based criterion that distinguishes long-lived association from fleeting contacts.

*Bound State*

A DNA/PAA configuration (frame) was considered contacting if:

i. The minimum heavy-atom distance $d_{min}$ between DNA and PAA was ≤ 0.6 nm:

$$d_{min}(t) = min_{i \in DNA, j \in PAA} \left\| r_i(t) - r_j(t) \right\| \quad (1)$$

ii. At least two hydrogen bonds ($N_{HB}$) existed between DNA and PAA molecules.

$$N_{HB}(t) \geq 2 \quad (2)$$

After defining the instantaneous contacting criteria, we treated the trajectory as a time series of a binary contact function:

$$\chi_{contact}(t) = \begin{cases} 1, & d_{min}(t) \leq 0.6 \ \ and \ N_{HB}(t) \geq 2 \\ 0, & otherwise \end{cases} \quad (3)$$

iii. To avoid including transient contacts, we further required persistence: a bound state was defined as any continuous time interval in which the above two criteria remained satisfied for at least $\tau_{\text{min}} = 100\text{ns}$.

We defined the final bound state indicator:

$$\chi_{bound}(t) = \begin{cases} 1, & if \ t \ lies \ inside \ any \ interval \ of \ length \ \geq \tau_{min} \\ 0, & otherwise \end{cases} \quad (4)$$

The bound fraction of a trajectory was calculated as the time average of this indicator:

$$f_{bound} = \frac{1}{T} \int_0^T \chi_{bound}(t) dt \quad (5)$$

where T is the total analyzed simulation time.

The 0.6 nm cutoff follows common practice in MD studies that quantify heavy-atom interfacial contacts using 0.5–0.6 nm thresholds[58–63]. This range is widely accepted because it captures first-shell, solvent-shared interactions beyond strict hydrogen bonds while excluding clearly solvent-separated configurations. The hydrogen-bond definition (donor–acceptor distance ≤ 0.35 nm; D–H···A angle ≥ 135°) is consistent with standard geometric criteria established in structural bioinformatics and MD analyses[64–69]. These cutoffs balance physical realism and statistical prevalence, they encompass the majority of energetically favorable hydrogen bonds while excluding weak, long-range or water-mediated contacts. The persistence window (100ns) was chosen to be well above the fast, transient processes that dominate polyelectrolyte environments, including hydrogen bonds and counterion dynamics, thereby filtering out ion-flicker and brief solvent-mediated encounters. Indeed hydrogen-bond lifetimes in bulk water are extremely short, typically on the order of ~0.7–1 ps in ab initio and MD simulations (e.g., ~0.78 ps in ab initio MD), with classical MD studies often yielding ~1 ps as a representative value[70,71]. These short lifetimes indicate that hydrogen bonds break and re-form on picosecond timescales, so a persistence threshold of 100 ns is many orders of magnitude longer than normal H-bond fluctuations. Counterion residence near DNA phosphates is usually tens to hundreds of picoseconds for monovalent ions and up to nanoseconds depending on groove or atom type[72–74]. Divalent ions such as $Ca^{2+}$ can form relatively long-lived phosphate-bridging contacts (e.g. ~98 ± 14 ns in DNA–lipid systems)[75], making 100 ns a conservative filter relative to these slower but still transient reorganizations. DNA conformational fluctuations, including bending and backbone dynamics, are rich below a few hundred nanoseconds; using a 100 ns requirement ensures that only multi–tens-of nanoseconds cooperative episodes, rather than frame-to-frame jitter, qualify as bound[76,77] Moreover, polyelectrolyte complexes are known to exhibit highly dynamic association with frequent exchange events; a 100 ns persistence criterion thus focuses analysis on sustained complexation rather than fleeting approach–separation cycles[78–80]. Short flexible polymers such as PAA are known to relax their conformations on nanosecond to few-tens-of-nanoseconds timescales[81,82]. Setting the persistence threshold at 100 ns therefore provides ample time for the chain to explore its accessible configurations and ensures that interactions classified as bound correspond to relaxed-state association rather than transient coil fluctuations.

*Calculation of Structural Parameters*

Using the final bound-state indicator $x_{bound}(t)$ and the corresponding bound fraction $f_{bound}$, we next analyzed DNA structural properties conditionally on whether frames were bound or unbound. In order to investigate the structural behavior of DNA in the presence of PAA, we calculated standard conformational descriptors, Root Mean Square Deviation (RMSD) and Radius of Gyration (Rg). These properties were evaluated with respect to the bound state classification described above, using the bound state indicator to distinguish between persistent bound and unbound states. For DNA-PAA simulations, structural properties were computed conditionally: averages were calculated over frames classified as bound to obtain bound state values, $\langle P\rangle_{bound}$, and separately over frames classified as unbound to obtain unbound state values, $\langle P\rangle_{unbound}$. Here, P stands for the structural property being measured; $\langle P\rangle_{bound}$ and $\langle P\rangle_{unbound}$ correspond to the time averages of the property P over bound and unbound frames, respectively. For DNA-alone and PAA-alone simulations, all frames were considered unbound, and only overall averages were calculated.

A time-averaged overall for each trajectory was then computed by combining bound and unbound contributions according to the fraction of time spent in the bound state $f_{bound}$(Equation 5):

$$\langle P\rangle_{overall} = f_{bound}\langle P\rangle_{bound} + (1 - f_{bound})\langle P\rangle_{unbound} \qquad 6$$

Replica averaging across multiple simulations (3 replicas) for each salt condition was performed, and the mean and standard error of the mean were reported to quantify variability. Our approach ensures that structural analyses reflect the dynamic association behavior of DNA with PAA, capturing both the effects of sustained complexation and intervals when the DNA is temporarily unbound.

*Identification of Ion Mediated DNA-PAA Contacts*

The interaction behavior between DNA and PAA chains in the presence of different salts was quantified by calculating the number of contacts between the polymer and specific regions of DNA. Contacts were analyzed separately for the backbone and for atoms defining the major and minor grooves, which are the two alternating wide and narrow grooves running along the outside of the double helix, caused by the uneven spacing of the sugar-phosphate backbones and based on atom selections that correspond to their chemical positions in the DNA double helix. The backbone contacts were identified from phosphate and sugar atoms, the major groove was defined by N7 and O6 atoms of purines and N4 and O4 atoms of pyrimidines, and the minor groove was defined by N3 atoms of purines and O2 atoms of pyrimidines. For each system and replica, the number of DNA-PAA contacts was computed as a function of simulation time using *gmx mindist* with a cutoff distance of 0.35 nm between heavy atoms. To exclude fleeting or electrostatically driven fly-by encounters that do not represent stable complexation, only bound frames were considered in the analysis. These bound intervals were identified from the persistence-based criterion described above, ensuring that contact statistics reflect sustained association events rather than transient proximity. Because the duration of bound intervals differed across replicas, the contact data within these intervals were resampled onto a uniform 1 ns time grid so that all replicas could be compared on the same temporal axis. The results from the three replicas per salt were then averaged to obtain the mean number of contacts as a function of time, while the standard deviation among replicas was used as an uncertainty measure and represented as shaded regions around the mean curves. Our procedure enabled a consistent comparison across different salts by minimizing fluctuations due to replica variability and focusing on bound configurations that represent long-lived DNA–PAA association.

To distinguish direct DNA-PAA contacts from ion-mediated associations, we separately quantified ion-mediated associations between DNA and PAA by detecting [DNA–ion–PAA] triplets. Direct contacts correspond to short-range interactions such as heavy-atom proximity and hydrogen bonding, which were already used in the bound-state definition. In contrast, ion-mediated binding occurs when a cation simultaneously interacts with DNA and PAA, forming a [DNA–ion–PAA] bridging configuration. A triplet was identified when a single cation simultaneously resided within 0.35 nm of at least one DNA atom belonging to a defined structural region (backbone, major groove, or minor groove) and within 0.35 nm of a PAA heavy atom. This cutoff corresponds to the first-shell coordination distances typically observed for divalent cations near phosphate and carboxylate oxygens, ensuring that only physically meaningful bridging events were considered[83].

For each system and replica, triplet counts were calculated for every frame using proximity data between ions and both molecular partners. Bound-state intervals were determined from the persistence-based criterion described above (including the 100ns requirement), and only frames belonging to these intervals were included in the analysis to exclude transient, non-associative encounters. The number of triplets within each bound interval was averaged to obtain a mean value per replica and region. Replica-averaged means and standard deviations were then computed for each salt, yielding salt-dependent measure of ion-mediated DNA–PAA association across distinct structural regions of the double helix.

*Interaction Energy Decomposition between DNA and PAA*

In the context of this study, the interaction between DNA and PAA is quantified through a short-range intermolecular interaction energy, denoted as $E_{total}$ . This quantity is defined as the sum of the short-range Lennard–Jones ($SR - E_{LJ}$) and short-range Coulomb interaction energies ($SR - E_{coul}$) between the two macromolecules. Energy groups corresponding to DNA and PAA were defined in the simulation input, and all reported interaction energies include only non-bonded interactions between atoms belonging to different groups. Intramolecular interactions within DNA and within PAA are therefore excluded by construction.

The total interaction energy in this context is defined as:

$$E_{total}(t) = E_{LJ,SR}(t) + E_{Coul,SR}(t) \qquad 7$$

where both contributions are evaluated as sums over all atom pairs $i \in$ DNA and $j \in$ PAA. The short-range Lennard Jones term $SR - E_{LJ}$ corresponds to direct van der Waals interaction between DNA and PAA atoms computed within the van der Waals cutoff distance $r_{vdW}$=1.0 nm. Electrostatic interactions were treated using the Particle Mesh Ewald method (PME). Within this framework, the Coulomb interaction is partitioned into a real-space short-range component and reciprocal space long-range component. The short-range Coulomb contribution $SR - E_{coul}(t)$ is calculated by direct summation of the Coulomb potential between DNA and PAA atom pairs separated by distances smaller than the real-space Coulomb cutoff $r_{Coulomb}$=1.0 nm. The long-range electrostatic contribution arising from the reciprocal-space part of the PME algorithm is fully included in the force evaluation and in the total electrostatic energy of the system but is not decomposed at the level of individual energy groups and is therefore not included in $E_{total}$ as defined here. As a result, $E_{total}$ represents the short-range intermolecular interaction energy between DNA and PAA and provides a direct measure of contact formation and dissociation between the two macromolecules. This definition of $E_{total}$ is used consistently throughout the manuscript and corresponds to the interaction energy profiles reported in Figure 1 and Figure S2.

## 3. RESULTS AND DISCUSSION

*Bound and Unbound States of the DNA-PAA Complexes*

To quantify the interaction between DNA and PAA, we analyzed the short-range intermolecular interaction energy between the two macromolecules, decomposed into short-range Coulombic and Lennard–Jones contributions, as defined in the Methodology section. The reported energies include only non-bonded interactions between DNA and PAA atoms and exclude intramolecular contributions as well as interactions with solvent and ions. The total interaction energy $E_{total}$, as defined in Equation 7, was used to monitor association and dissociation events and to characterize the salt-dependent interaction behavior of the DNA–PAA system.

The interaction energy decomposition in Figure S2 shows that the short-range Coulombic contribution SR-$E_{Coul}$ remains mostly positive in all salt environments. This indicates that, at short distances and considering only the real-space contribution, DNA-PAA electrostatic interactions are generally repulsive, while transient values close to zero or weakly negative are occasionally observed. These fluctuations arise from the heterogeneous atomic charge distributions of DNA and partially deprotonated PAA, which allow locally favorable short-range electrostatic contacts, such as those involving base or sugar moieties and PAA carboxylate or hydroxyl groups, to occur during close, hydrogen-bonded configurations.  The magnitude and fluctuations of this repulsive contribution vary among salts. In NaCl, the SR-$E_{Coul}$ term stays moderately positive throughout the trajectory. In $MgCl_2$, it approaches values closer to zero for long intervals, reflecting reduced repulsion during those periods. In $CaCl_2$, the SR-$E_{Coul}$ term remains positive but shows broader fluctuations than in NaCl, with intervals where the repulsion is weaker. The short-range Lennard-Jones contribution SR- $E_{LJ}$ is negative in all salts, reflecting favorable van der Waals and contact interactions. The depth and persistence of the SR-$E_{LJ}$ wells differ between salt conditions. In NaCl, SR- $E_{LJ}$ energies remain consistently negative across long segments of the trajectory. In $MgCl_2$, SR- $E_{LJ}$ energies stay near zero for extended intervals and become negative only during shorter contact events. In $CaCl_2$, SR-$E_{LJ}$ energies are negative for most of the trajectory and reach deeper minima in certain intervals. The combination of these repulsive SR-$E_{Coul}$ and attractive SR-$E_{LJ}$ components produces the total interaction energy, and the changes in these profiles indicate how often and how strongly DNA and PAA come into contact under each salt condition. These energetic variations directly reflect the fluctuating association behavior of the two polyelectrolytes.

DNA and PAA, both highly charged polyelectrolytes, exhibit dynamic, intermittent association. The differences observed in the energy profiles of simulation replicas for the same salt type (Figure S2), as well as the variations in energy within a single replica over the course of the simulation, reflect the fluctuating association behavior. Figure 1 shows representative single molecule DNA-PAA interaction energy profiles of a simulation replica for different salt types. The specific replicas shown in Figure 1 were selected because, especially in the NaCl and $MgCl_2$ simulations, they display clear alternations between weakly interacting ($E \approx 0$) and strongly interacting periods, which makes them the most illustrative examples of the association–dissociation behavior analyzed in this study. As can be seen in Figure 1b, in $MgCl_2$ environment the interaction energy between DNA and PAA is approximately zero around 650–800 ns, indicating no interaction between the two structures, followed by noticeable differences in interaction energy around 900–1000 ns. The observed fluctuations therefore do not imply the absence of equilibrium but instead indicate that the DNA-PAA system does not settle into a single, long-lived bound configuration. Rather, it continually explores associated and dissociated states, consistent with prior reports that polyelectrolyte complexes can occupy multiple metastable configurations and exhibit dynamic, partially associated behavior[84]. Zero or near-zero interaction energy segments correspond to time intervals in which the DNA–PAA pair does not satisfy the distance and hydrogen-bond requirements of persistent binding, so these regions align directly with unbound periods in our state classification.

Based on these observations, in this study, instead of assuming a conventional binding equilibrium-like state in MD simulations, we focused on the bound and unbound states defined in Section 2.2. and performed all structural calculations with respect to these states. To illustrate this approach, Figure S3, S4 and S5 presents the bound-state detection for the three salt environments (NaCl, $MgCl_2$, $CaCl_2$) of 3 replicas using the distance and hydrogen-bond criteria.

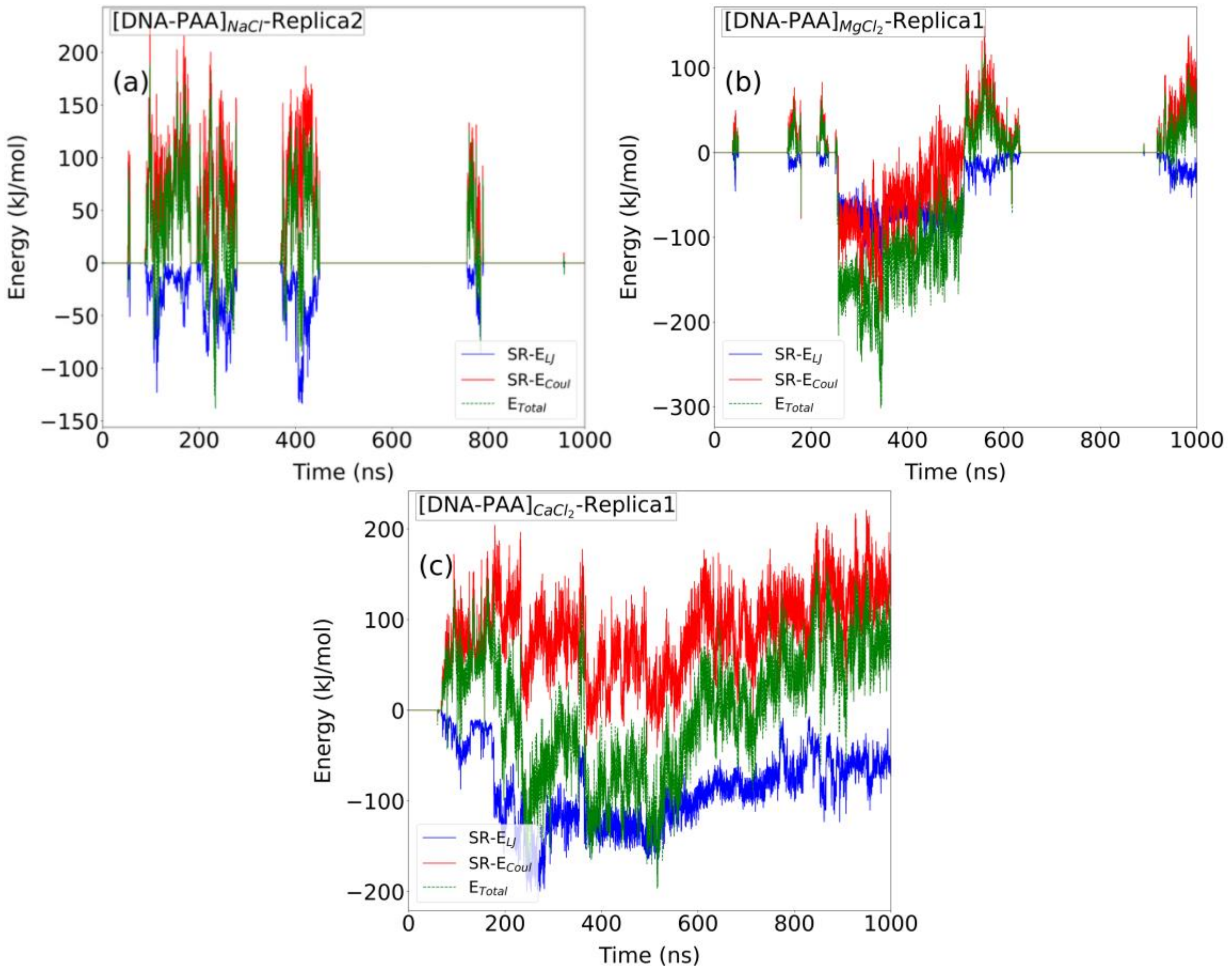

**Figure 1.** Time evolution of DNA–PAA interaction energies Total Energy ($E_{Total}$), SR-LJ (SR-$E_{LJ}$) and SR-Coulombic (SR-$E_{Coul}$) contributions in different salt environments: (a) NaCl, (b) $MgCl_2$, (c) $CaCl_2$.

The bound fraction analysis ($f_{bound}$) reveals clear salt-dependent binding behavior, as shown in Figures S3–S5 and summarized in Table 2, with the calculation procedure described in Section 2. 2. In NaCl (Figure S3), the binding profile varies strongly across replicas: some trajectories display long periods of dissociation with intermittent contacts (Replica 2), while others form early and persistent complexes (Replica1 and Replica 3). In $MgCl_2$ (Figure S4), binding becomes generally more stable but still shows intermittent detachment in certain replicas before settling into lasting associations (Replica 1 and Replica 2). In contrast, $CaCl_2$ (Figure S5) leads to consistently strong and continuous binding in all trajectories, with the DNA–PAA complex remaining associated for nearly the entire simulation. The time-weighted bound fractions reported in Table 2 quantify these trends, showing that $Ca^{2+}$ markedly stabilizes the complex, whereas $Mg^{2+}$ and $Na^{+}$ permit more dynamic and fluctuating binding. This establishes the overall trend in salt-dependent complex stability, which is further examined at the molecular level through the radial distribution of cations around DNA and PAA.

**Table 2.** Time-weighted bound fraction $f_{bound}$ for each replica and mean ± standard deviation for the three salt environments.

| Salt | Replica 1 | Replica 2 | Replica 3 | Mean ± SEM |
|---|---|---|---|---|
| NaCl | 0.64 | 0.0 | 0.94 | 0.53 ± 0.28 |
| $MgCl_2$ | 0.34 | 0.28 | 0.94 | 0.51 ± 0.21 |
| $CaCl_2$ | 0.93 | 0.76 | 0.99 | 0.89 ± 0.06 |

For each system we further computed radial distribution functions g(r) between the cations and the phosphate oxygens of DNA, as well as between the cations and the carboxylate oxygens of PAA, averaging the results over the three simulation replicas for each salt. Results are showing in Figure 2, where the RDFs are plotted on a log10 scale to improve the visibility of peak height differences among the ions. The ion–oxygen radial distribution functions reveal clear differences in the way the three cations associate with DNA and PAA. $Ca^{2+}$ shows a very sharp and intense first-shell maximum near 0.25–0.30 nm for both the DNA phosphate oxygens and the PAA carboxylate oxygens, indicating frequent direct coordination. $Na^{+}$ also exhibits a distinct first peak at a similar position but with lower amplitude. In contrast, $Mg^{2+}$ displays only a small inner-sphere feature and its main maximum appears farther out, around 0.38–0.45 nm for both polyelectrolytes. Monovalent $Na^{+}$, and divalent $Mg^{2+}$ and $Ca^{2+}$ differ in their hydration and coordination behaviour, and these differences offer a plausible basis for the salt-dependent binding trends. $Na^+$ has a lower charge density and a more flexible hydration shell, which allows frequent water exchange and relatively accessible coordination to oxygens in the surrounding media[85]. $Mg^{2+}$ is characterized by a small ionic radius and high charge density, which leads to a tightly held first hydration shell and slower water exchange[86]. $Ca^{2+}$ has a larger ionic radius, a more labile hydration environment and a coordination number that fluctuates more than that of $Mg^{2+}$, which permits more frequent direct interactions with oxygens[87].

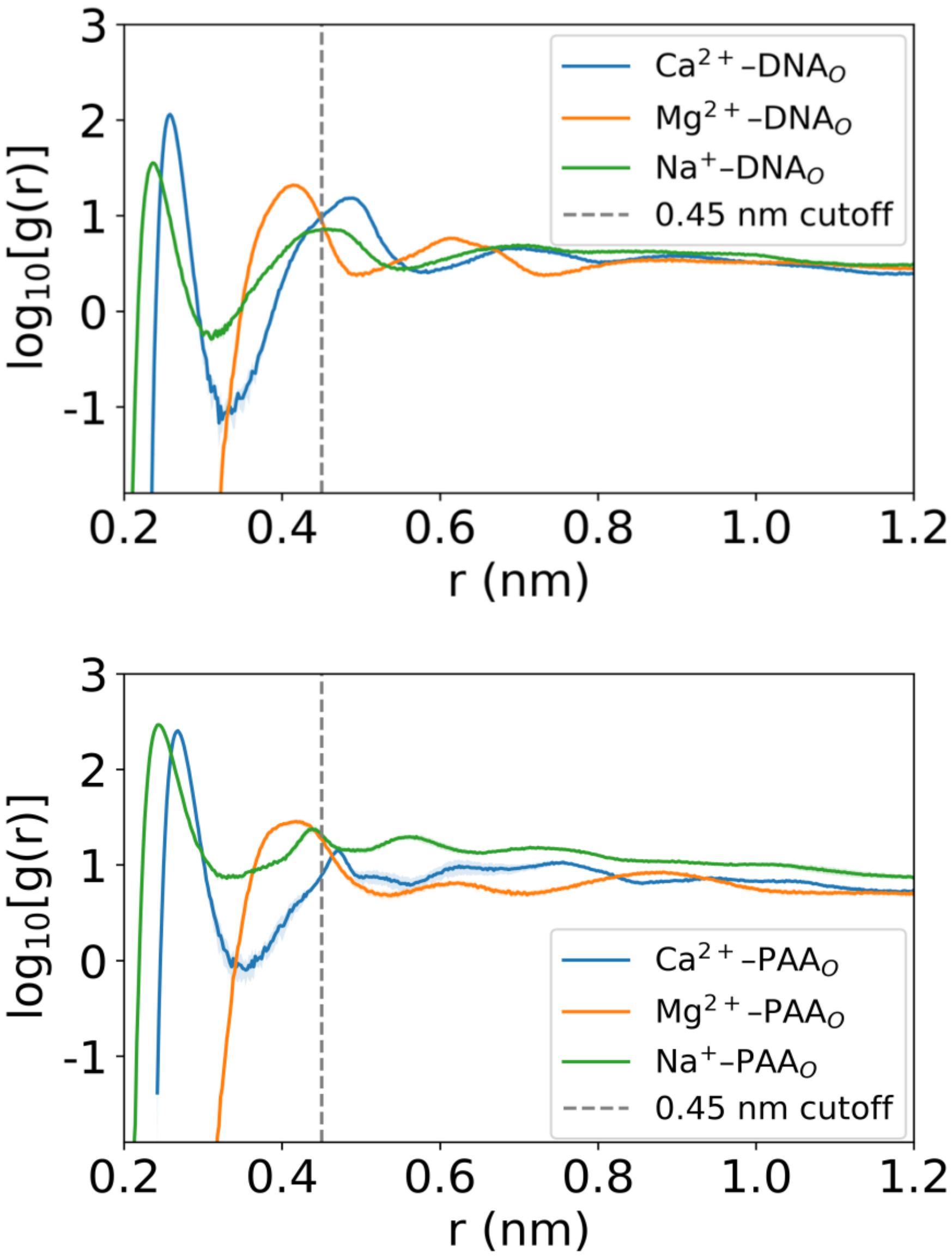

**Figure 2.** Radial distribution functions g(r) of $Ca^{2+}$, $Mg^{2+}$ and $Na^{+}$ around the phosphate oxygens of DNA (top) and the carboxylate oxygens of PAA (bottom), averaged over three simulation replicas for each salt. The dashed line marks the 0.45 nm distance.

Because $Ca^{2+}$, $Mg^{2+}$ and $Na^{+}$ show clearly different first-shell coordination patterns in the RDF analysis (Figure 2), we next examined whether these differences translate into direct cross-linking of the two polyelectrolytes. We quantified bridging events, defined as a single ion simultaneously contacting a DNA phosphate oxygen and a PAA carboxylate oxygen within 0.45 nm. The cutoff distance is consistent with the locations of the first minima in the computed RDFs (Figure 2) and with commonly used distances for ion–oxygen coordination in nucleic acids reported in the literature, where inner-sphere binding of divalent cations is typically found at 0.25–0.30 nm and the first solvation shell extends to about 0.4–0.45 nm[88–90].

Table 3 summarizes the mean number of bridges per bound frame and the fraction of bound-state time containing at least one bridge. The percentage of bound frames containing at least one ion bridge was computed using the resampled 1 ns trajectory, meaning that each time point corresponds to a 1 ns interval within the bound-state window. $Ca^{2+}$ not only forms the largest number of bridges but also maintains at least one cross-link during ~72 % of the time the complex is bound, compared with ~44 % for $Mg^{2+}$ and only ~10 % for $Na^{+}$. This demonstrates that $Ca^{2+}$ can effectively lock the two polyelectrolytes together through persistent direct coordination, while $Mg^{2+}$ provides weaker, less frequent bridging and $Na^{+}$ contributes minimally. The substantial bound fractions observed in NaCl and $MgCl_2$ (Table 2) therefore arise mainly from electrostatic screening: both salts reduce the backbone repulsion enough for DNA and PAA to associate via other interactions such as hydrogen bonding and van der Waals contacts. In contrast, $Ca^{2+}$ not only screens but also acts as a true molecular cross-linker, stabilizing the complex through frequent and long-lived bridges. Differences in ion mobility provide additional insight into this behaviour, as shown by the replica averaged Mean Square Displacement (MSD) profiles in Figure S6 calculated using equation S1. The mean squared displacement curves show a small downward curvature at long times. This behavior is a known finite-size artifact in periodic simulations, arising from hydrodynamic correlations and the coupling to periodic images, which can lead to apparent deviations from a strictly linear long-time MSD even for simple diffusive motion[91]. Reliable diffusivity estimates are therefore obtained from the time window where the MSD is well described by a linear regime, and we base the comparison of ion mobility on the long-time slope extracted from this linear region rather than on the slightly sub-linear tail. Since the present analysis is used for qualitative comparison of ion mobility through the relative slopes, the application of finite-size diffusion corrections was not required.

$Mg^{2+}$ shows the lowest mobility among the three ions, in line with previous work indicating a structurally and dynamically rigid first hydration shell with stable sixfold octahedral coordination and slow water exchange compared with $Ca^{2+}$[92]. $Ca^{2+}$ exhibits higher mobility, consistent with its more labile and structurally flexible hydration environment, while $Na^{+}$ diffuses the fastest, a behaviour that matches its weak hydration and transient interactions with the two polyelectrolytes.

In Figure 3, representative simulation snapshots illustrate how $Ca^{2+}$ and $Mg^{2+}$ ions mediate these cross-links by simultaneously coordinating phosphate oxygens on DNA and carboxylate groups on PAA. In these examples, one or more $Ca^{2+}$ or $Mg^{2+}$ ions sit directly between the two polymers, forming stable coordination sites that lock the chains together and exemplify the bridging mechanism quantified in our analysis.

**Table 3.** Average number of ion-mediated DNA–PAA bridges and fraction of bound-state time containing at least one bridge for each salt, averaged over three simulation replicas.

| **Ion** | **Mean bridges in bound frames** | **% bound frames with ≥1 bridge** |
|---|---|---|
| $Ca^{+2}$ | 2.17 ± 1.33 | 72.1 % |
| $Mg^{+2}$ | 1.08 ± 1.11 | 44.2 % |
| $Na^{+}$ | 0.15 ± 0.11 | 9.8 % |

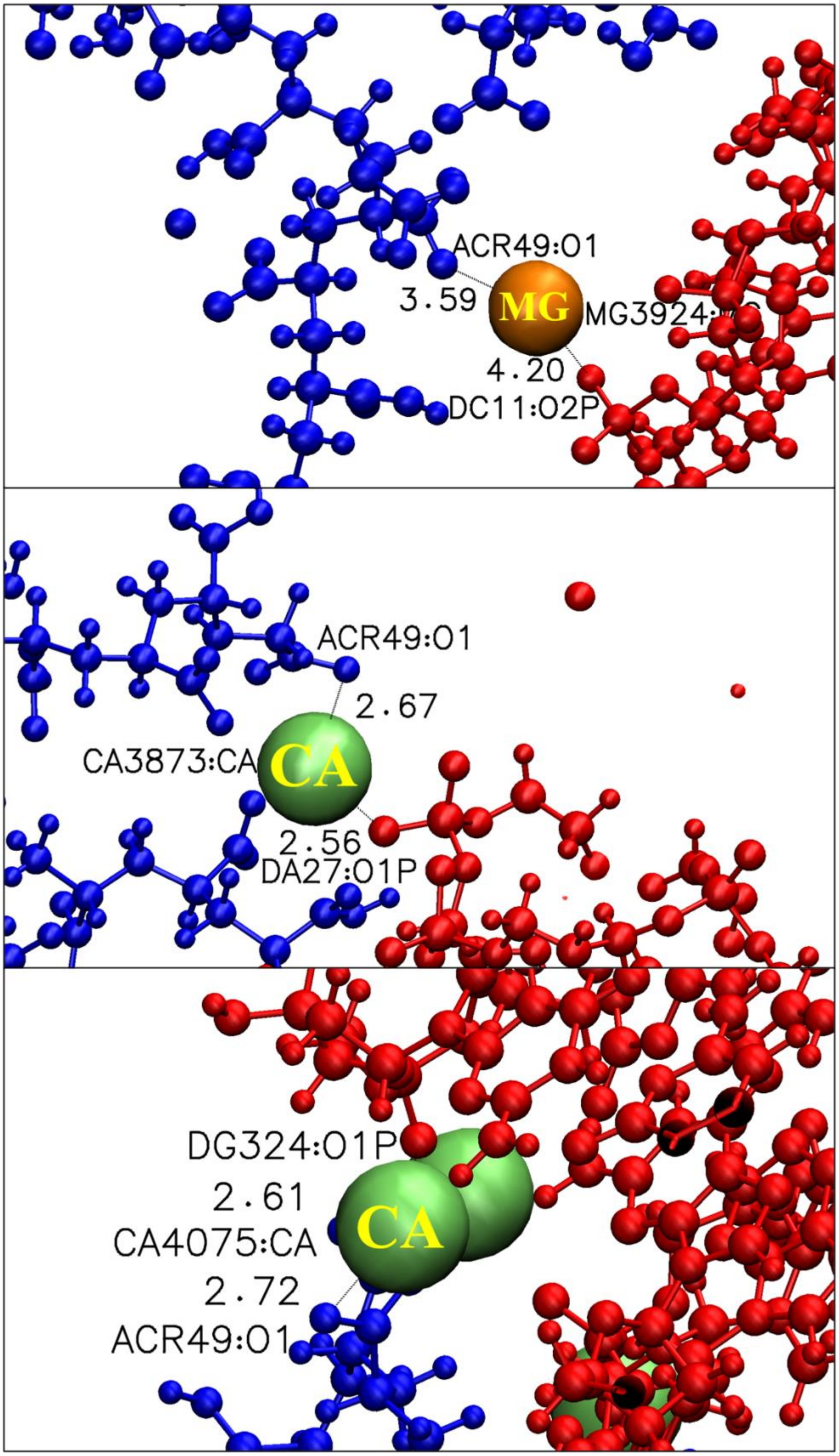

**Figure 3.** Representative simulation snapshots showing $Ca^{2+}$ (green spheres) and $Mg^{2+}$ (orange spheres) bridging DNA phosphate groups (red) and PAA carboxylate groups (blue).

While the cation-mediated bridging analysis provides insight into how multivalent ions stabilize the DNA–PAA association, it does not directly reveal how the two polyelectrolytes approach and organize relative to one another. To probe the intrinsic spatial correlations between the polymers themselves, we next computed radial distribution functions of PAA oxygen atoms around the phosphate oxygens of DNA. These RDFs capture how the carboxylate-rich PAA chains distribute with respect to the negatively charged DNA backbone under the different salt conditions and it complements the cation–oxygen analyses by focusing on direct polyelectrolyte–polyelectrolyte structuring rather than ion-mediated interactions.

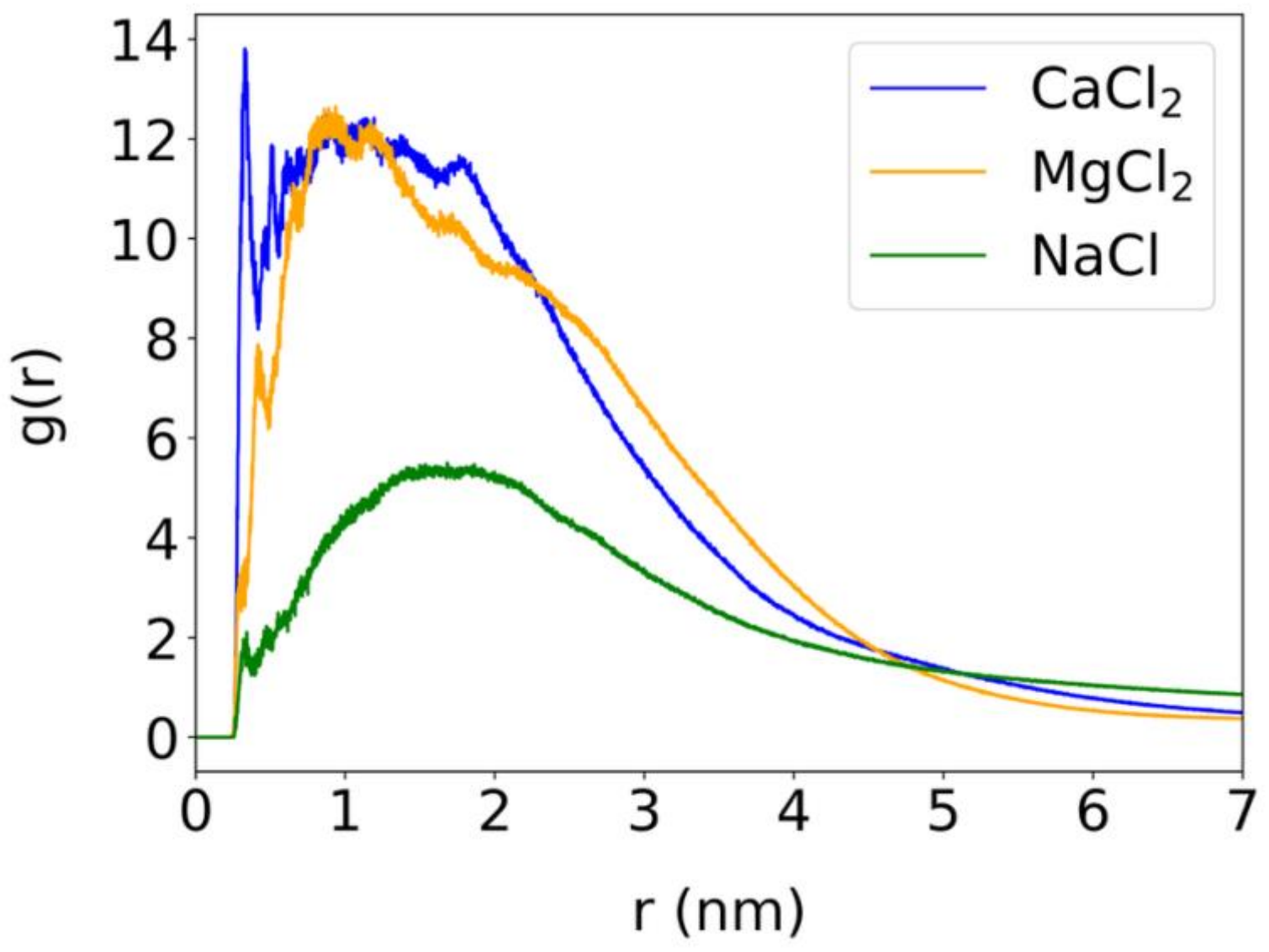

**Figure 4.** Radial distribution of PAA oxygen atoms around DNA phosphate oxygens in different salt environments.

The PAA–DNA radial distribution functions shown in Figure 4 reveal a clear salt dependence in polyelectrolyte–polyelectrolyte structuring. Complexes in $CaCl_2$ display a sharp first maximum at ~0.3–0.4 nm, indicating tight, well-defined contact of PAA carboxylates with DNA phosphates. $MgCl_2$ shows a weaker and broader first peak, consistent with more transient association. NaCl yields only a shallow, diffuse maximum, reflecting loose screening-driven contacts. This behavior agrees with the cation coordination and bridging analysis, which showed that $Ca^{2+}$ acts as a strong cross-linker, $Mg^{2+}$ provides moderate stabilization, and $Na^{+}$ contributes mainly through electrostatic screening without persistent bridges. This difference between $Ca^{2+}$ and $Mg^{2+}$ follows directly from their distinct coordination environments identified in the RDF analysis, where $Ca^{2+}$ displays a more accessible and flexible first-shell arrangement than the more tightly hydrated $Mg^{2+}$.

*Structural Properties of DNA*

Having established how long the complexes remain associated under each salt condition, we next asked how DNA's own conformation responds to binding. To separate intrinsic DNA fluctuations from complexation effects, we calculated RMSD and Rg of DNA for each trajectory by classifying frames into the bound and unbound intervals previously defined. This approach allows two direct comparisons: (i) within a given simulation, DNA when complexed versus DNA when not complexed; (ii) unbound DNA in the presence of PAA versus DNA simulated alone in the same salt. For completeness, we also report a time-weighted overall value per trajectory, in which the bound and unbound contributions are combined according to the measured $f_{bound}$. RMSD values from the three independent replicas were averaged and are reported as mean ± SEM in Table 4, while the corresponding Rg values are summarized in Table 5.

**Table 4.** State-resolved heavy-atom RMSD (nm) by salt for DNA alone and DNA–PAA systems.

| Salt | $RMSD_{bound}$ (nm) | $RMSD_{unbound}$ (nm) | $RMSD_{overall}$ (nm) | $RMSD_{[DNA]alone}$ (nm) |
|---|---|---|---|---|
| NaCl | 1.2 ± 0.2 | 1.2 ± 0.2 | 1.2 ± 0.08 | 1.0 ± 0.001 |
| MgCl2 | 0.9 ± 0.1 | 1.0 ± 0.1 | 0.9 ± 0.1 | 1.0 ± 0.002 |
| CaCl2 | 1.1 ± 0.1 | 0.8 ± 0.2 | 1.1 ± 0.1 | 1.04 ± 0.002 |

Overall, RMSD values show little difference between bound and unbound DNA in NaCl and $MgCl_2$, and their values are essentially the same as DNA simulated alone, indicating that these salts and PAA do not substantially perturb the DNA backbone. The only clear effect is observed in $CaCl_2$, where DNA in the bound state exhibits a higher RMSD (≈ 1.1 nm) compared with both its unbound segments (≈ 0.8 nm) and DNA simulated alone (≈ 1.0 nm). This suggests that $Ca^{2+}$-mediated cross-linking perturbs DNA conformation more strongly than $Na^{+}$ or $Mg^{2+}$, consistent with the bridging and RDF analyses.

In parallel, we examined the conformational stability of the PAA by analyzing its heavy-atom RMSD using the same bound–unbound classification as shown in Table S3. Across all salt conditions, the RMSD of PAA remains close to the values observed for PAA simulated alone, indicating that complexation with DNA does not induce large-scale structural rearrangements of the polymer backbone. In NaCl and $MgCl_2$, the bound and unbound RMSD values of PAA differ only marginally and remain within statistical uncertainty of the PAA-alone reference. In $CaCl_2$, a modest increase in PAA RMSD is observed in the bound state relative to the unbound segments, reflecting localized conformational adjustments associated with $Ca^{2+}$-mediated bridging rather than global backbone deformation. Overall, results shown in Table S3 indicate that DNA–PAA association primarily perturbs DNA conformation under $CaCl_2$ conditions, while the PAA backbone retains its intrinsic flexibility across all salts.

Rg shows minimal state dependence in NaCl and only a small reduction upon binding in $MgCl_2$, both remaining close to the DNA-alone reference. In $CaCl_2$, however, a marked state effect emerges. When complexed with PAA, DNA exhibits a larger Rg than both its unbound segments and the DNA-alone control, whereas the unbound periods in $CaCl_2$ are relatively compact. This pattern is consistent with $Ca^{2+}$ mediated cross-linking that stabilizes bound conformations with a more extended global size, while $Ca^{2+}$ screening in the absence of persistent complexation permits a more compact coil. Also bridging analysis indicates that several adjacent backbone phosphates participate in the interaction (Figure 3). This shows that the complexation involves an extended portion of the DNA rather than a single localized region, which explains why the bound-state DNA adopts a more enlarged conformation in $CaCl_2$.

**Table 5.** State-resolved heavy-atom Rg (nm) by salt for DNA alone and DNA–PAA systems.

| **Salt** | **$Rg_{bound}$ (nm)** | **$Rg_{unbound}$ (nm)** | **$Rg_{overall}$ (nm)** | **$Rg_{[DNA]alone}$ (nm)** |
|---|---|---|---|---|
| NaCl | 2.9 ± 0.12 | 2.9 ± 0.12 | 2.9 ± 0.05 | 3.1 ± 0.17 |
| MgCl2 | 2.7 ± 0.11 | 2.9 ± 0.08 | 2.8 ± 0.08 | 2.8 ± 0.13 |
| CaCl2 | 3.0 ± 0.03 | 2.5 ± 0.15 | 2.9 ± 0.03 | 2.6 ± 0.07 |

The ordering of Rg for the DNA-alone systems (NaCl > $MgCl_2$ > $CaCl_2$) reflects well-known ion-specific effects on DNA structure. Monovalent $Na^+$ maintains stronger electrostatic repulsion along the DNA backbone, leading to a slightly more expanded conformation in solution. In contrast, divalent ions reduce these repulsions more effectively, and the extent of this reduction depends on the ion's hydration properties and coordination behavior. $Mg^{2+}$, with its small ionic radius and tightly bound first hydration shell, interacts with DNA mainly through outer-sphere coordination, producing a moderate degree of compaction. $Ca^{2+}$, which possesses a more labile hydration shell and can engage in more frequent direct contacts with phosphate oxygens, screens the backbone charges more efficiently and therefore yields the most compact DNA conformation. These trends are consistent with experimental[93] and theoretical studies[94] that report ion-dependent modulation of DNA stiffness and electrostatic screening, supporting the interpretation that the Rg differences observed here arise from intrinsic physical differences among $Na^+$, $Mg^{2+}$, and $Ca^{2+}$ in their interactions with the DNA double helix.

*Complexation Shape Analysis*

Although the complexation between the two molecules does not significantly alter the global structural descriptors of DNA such as RMSD and Rg, subtle rearrangements in the overall mass distribution and orientation may still occur. These differences can be captured through the analysis of the principal moments of inertia and their relative orientations, which provide deeper insight into the topology of the complex. The principal moments of inertia describe how the molecular mass is distributed around three orthogonal axes and therefore indicate how extended the structure is along each direction. Larger values reflect a broader spread of mass, so comparing $I_1$, $I_2$, and $I_3$ allows one to assess changes in molecular shape. To investigate these rearrangements, the principal moments of inertia ($I_1$, $I_2$, and $I_3$) of both DNA and PAA were calculated separately for the bound and unbound states in all replicas and salt conditions. The complete set of inertia values for all replicas is provided in Tables S1 and S2. These values are the mechanical principal moments of inertia obtained from diagonalization of the mass-weighted inertia tensor. To obtain representative averages, the moments were then averaged over three replicas for each salt environment, and their corresponding standard deviations were calculated. Our approach allows comparison of how complexation and ionic conditions influence the overall mass distribution of each molecule.

Subsequently, the relative orientations between the principal axes of DNA and PAA were examined to evaluate the geometrical organization of the complex.

Figure 5 presents the mean principal moments of inertia ($I_1$, $I_2$, and $I_3$) of DNA and PAA for the bound and unbound states across all salt conditions, averaged over three replicas in each ionic environment. In $CaCl_2$, all inertia components of DNA increase upon binding, indicating overall expansion and a more extended mass distribution within the complex. In $MgCl_2$, all I values decrease, showing that DNA becomes more compact when complexed, whereas NaCl produces only minor differences between states. For PAA, the moments of inertia decrease in $CaCl_2$ and NaCl upon binding, revealing compaction of the polymer within the complex, while $MgCl_2$ shows little change. These structural responses, where DNA expands and PAA becomes more compact in $CaCl_2$, reflect the strong cross-linking and alignment induced by $Ca^{2+}$ compared to the more flexible association observed under $Mg^{2+}$ and $Na^+$ conditions.

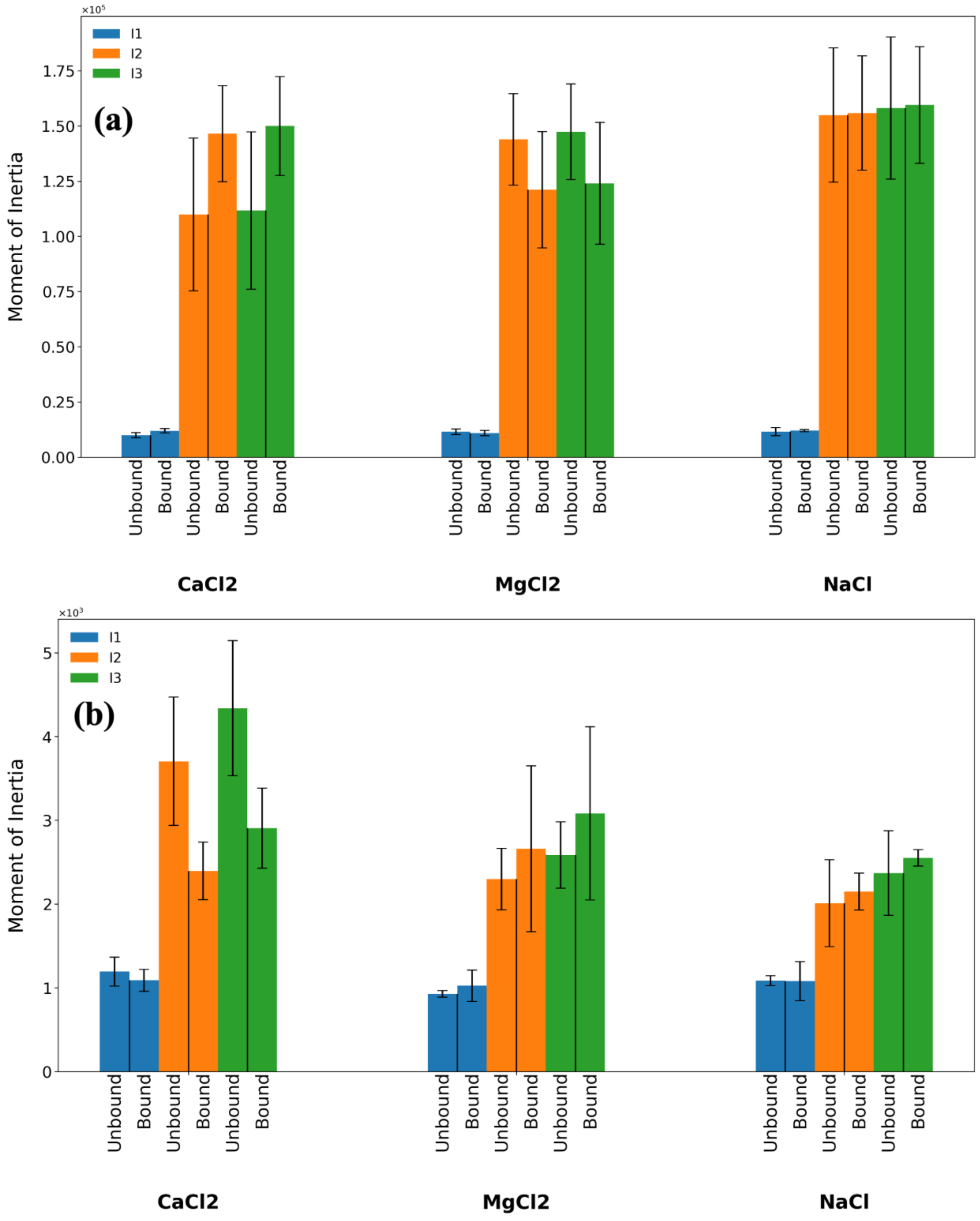

**Figure 5.** Principal inertia moments for DNA(a) and PAA (b) across salts in bound and unbound states averaged over three replicas.

The relative orientation between DNA and PAA was analyzed to quantify the structural organization of the complexes under different salt conditions. For each trajectory, the principal moments of inertia and the corresponding principal axes were extracted for both DNA and PAA using inertia tensor decomposition, where the first principal axis of DNA represents its molecular long axis and the third principal axis of PAA corresponds to its molecular short axis (Table S1, Table S2). The time series of these axes were synchronized by matching time stamps to ensure frame-wise correspondence between the two molecules. For each synchronized frame, the inter-axis angle $\theta$ was computed as $\theta = arccos(|u \cdot v|)$, where u and v denote the unit vectors of the DNA and PAA principal axes, respectively. The bound and unbound configurations were separated using the binding windows defined for each replica, and the corresponding $\cos\theta$ values were collected and combined across replicas for each salt. The normalized probability density $p(\cos\theta)$ was computed over the range 0–1. Using $\cos\theta$ removes the geometric bias inherent in angle space and provides a uniform reference for uncorrelated orientations, which allows a clearer comparison of orientational organization in NaCl, $MgCl_2$ and $CaCl_2$ solutions. Representative configurations corresponding to characteristic $\cos\theta$ values for each salt are shown in Figure 7, illustrating the relative alignment between the DNA long axis and the PAA short axis in the three ionic environments.

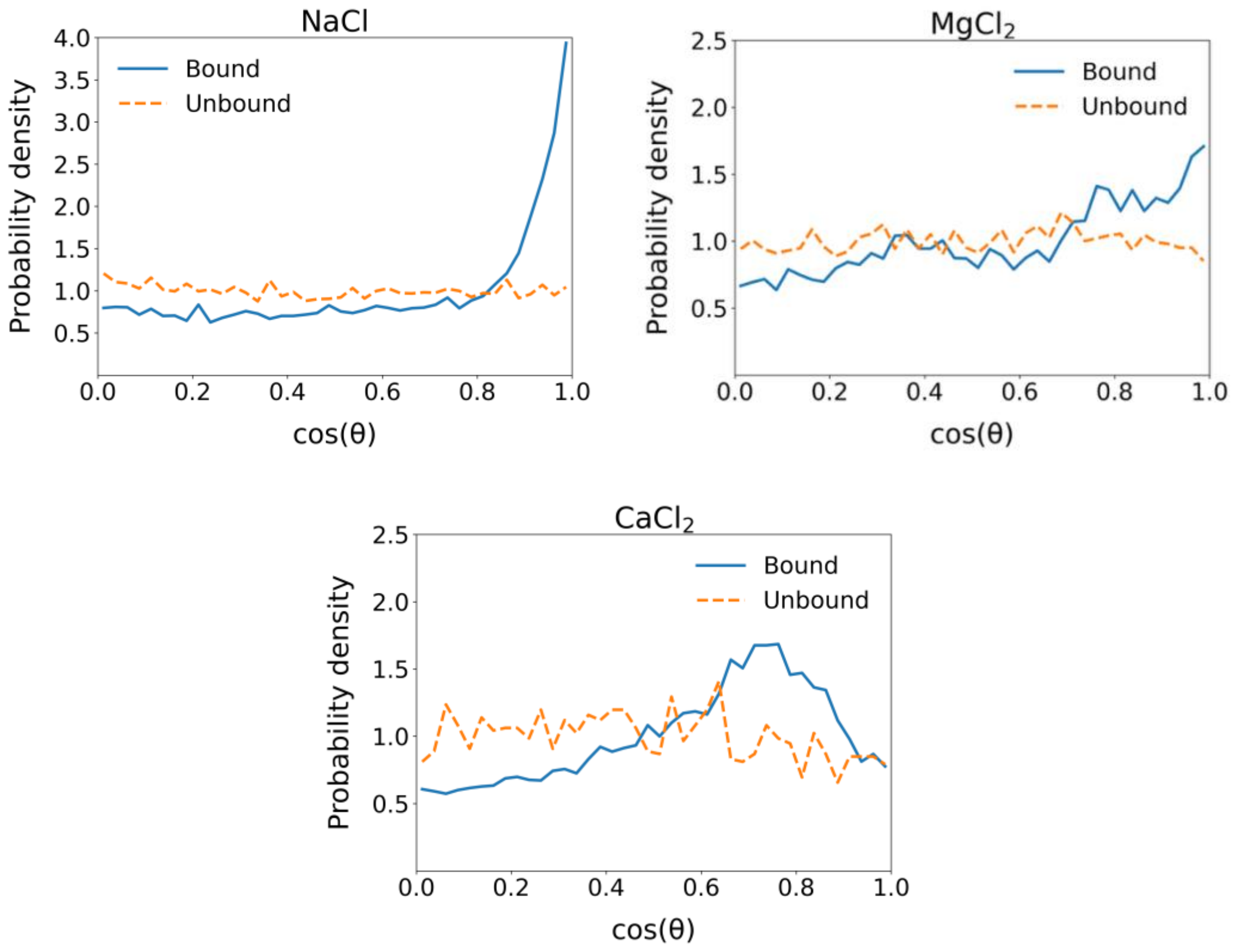

**Figure 6.** Normalized probability density functions of cosθ between the DNA long axis and the PAA short axis for bound and unbound configurations in NaCl, $MgCl_2$, and $CaCl_2$ solutions.

As shown in Figure 6, the probability distributions p(cosθ) for the DNA and PAA principal axes in NaCl, $MgCl_2$ and $CaCl_2$ solutions reveal clear ion-dependent trends in orientational organization. In NaCl, the unbound configurations remain nearly uniform across the entire cosθ range, consistent with largely uncorrelated orientations between the two molecules. Upon complexation, the distribution develops a strong rise toward cosθ approaching 1, indicating an increased population of configurations in which the DNA and PAA axes become nearly parallel. The gradual increase across intermediate cosθ values reflects a broad and dynamic orientational behavior under the weaker electrostatic screening of monovalent $Na^+$ ions. In $MgCl_2$, the bound state shows a more progressive enhancement toward higher cosθ values, with a smoother but more pronounced increase relative to NaCl. This behavior suggests that $Mg^{2+}$ promotes partial orientational ordering, consistent with transient bridging interactions that intermittently align the molecular axes. The effect becomes more distinct in $CaCl_2$, where the bound distribution exhibits a well-defined maximum around cosθ ≈ 0.7–0.8, indicating a more stable and persistent alignment of the DNA and PAA axes. This enhanced orientational ordering reflects the stronger cross-linking capacity of $Ca^{2+}$ ions. Across all salts, the unbound distributions remain comparatively flat, while the bound distributions shift progressively toward higher cosθ values from NaCl to $CaCl_2$, demonstrating an increase in orientational order with ion valency.

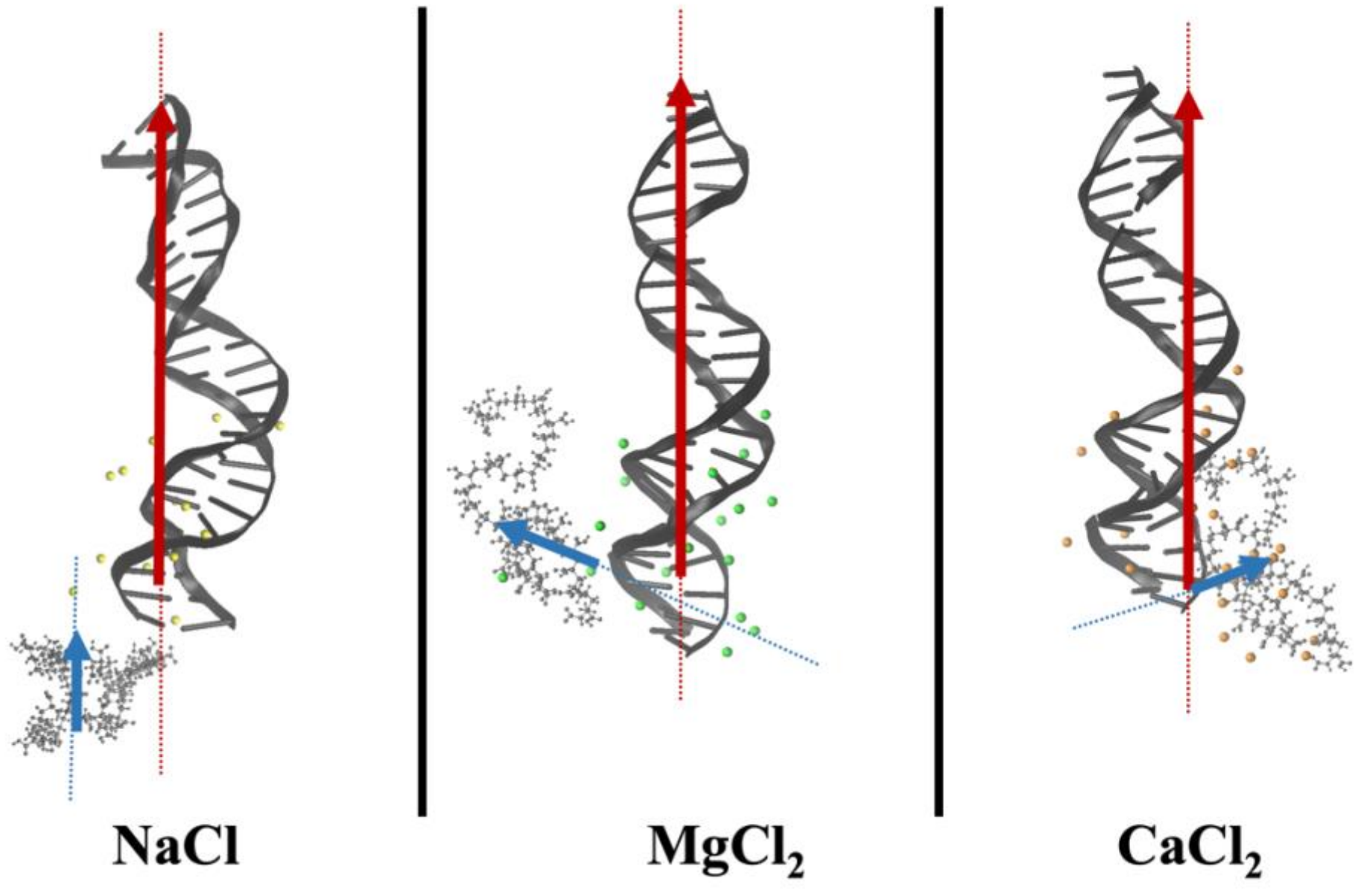

**Figure 7.** Representative bound-state configurations showing the orientation between the DNA long axis and the PAA short axis for NaCl, $MgCl_2$, and $CaCl_2$.

*Backbone and Groove Binding*

The spatial organization of DNA–PAA interactions was analyzed by quantifying contacts between DNA backbone and groove regions and PAA carboxyl atoms across different salt conditions. Each region of DNA was defined using chemically distinct atom sets, separating phosphate based interactions along the backbone from potential groove occupation. The evolution of contacts during bound intervals provides a detailed view of how the polymer engages with DNA and how ion valency modulates the balance between backbone attachment and groove association. The computational procedure for contact analysis is detailed in the Methods Section 2.2. Figure S8 presents the evolution of DNA–PAA contacts within the backbone, major, and minor grooves during the bound states across different salt environments. The comparison reveals distinct salt-dependent interaction modes. In NaCl, contact formation remains sparse and highly fluctuating for all DNA regions, indicating weak electrostatic screening and frequent polymer detachment. The few contacts that do occur are largely confined to the phosphate backbone, reflecting a predominance of long-range Coulombic attraction without stable bridging. In $MgCl_2$, contact density increases substantially, with both backbone and groove atoms participating more consistently in the interaction. This enhancement suggests that $Mg^{2+}$ ions partially compensate for the electrostatic repulsion between the negatively charged DNA and PAA, allowing the polymer to approach more closely and occasionally penetrate the grooves. The transient but repeated groove associations point to dynamic, short-lived binding mediated by partially coordinated $Mg^{2+}$ ions. In $CaCl_2$, the overall contact level is significantly higher, dominated by persistent backbone association. The polymer remains in continuous proximity to the DNA throughout the simulation, consistent with strong electrostatic screening and cation-bridged stabilization. Groove contacts are also present but occur much less frequently than in $MgCl_2$, indicating that $Ca^{2+}$ favors compact, backbone-bound configurations rather than deeply inserted groove binding. These trends demonstrate that increasing ion valency strengthens the overall DNA–PAA association by enhancing charge compensation and stabilizing the backbone interface. Yet, the simple contact metric does not indicate whether the observed interactions are directly mediated by ions or reflect independent electrostatic attractions. Therefore, to explicitly identify and quantify cation-bridged binding events, we next computed the number of [DNA–ion–PAA] triplets within each DNA region.

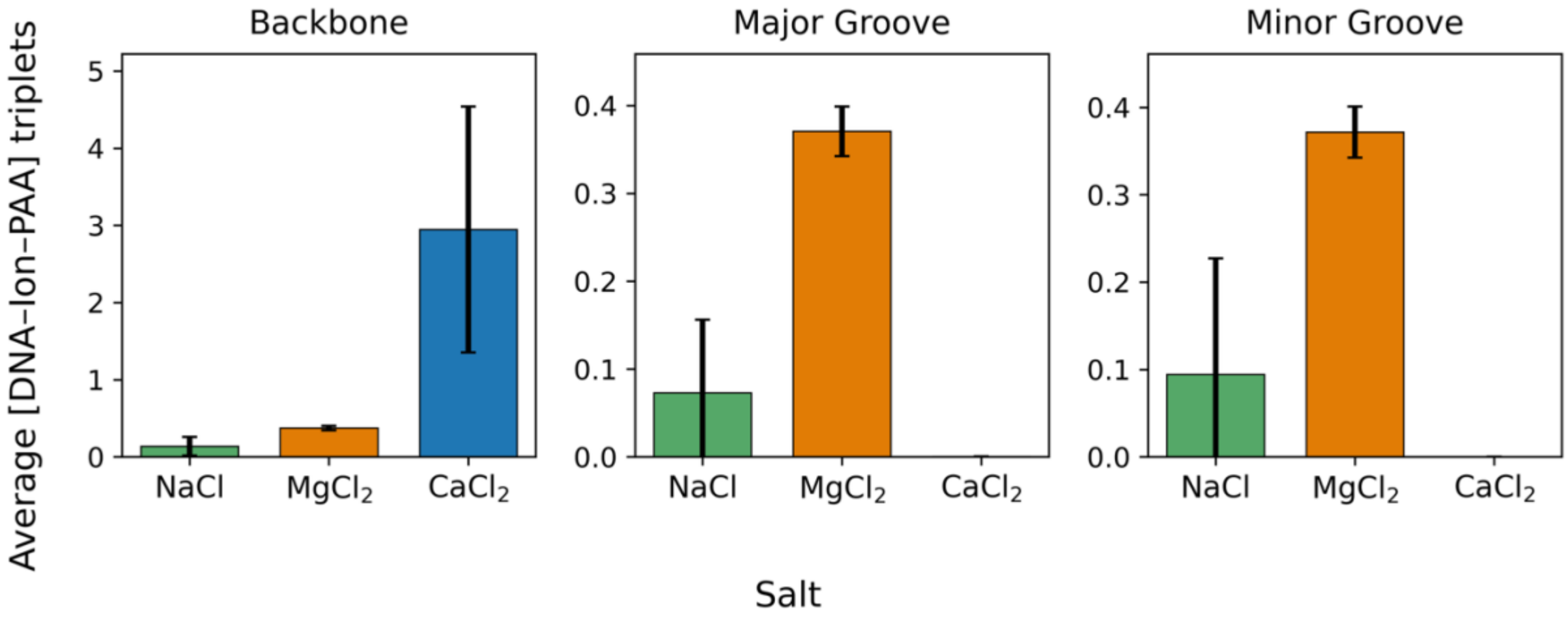

**Figure 8.** Average number of [DNA–ion–PAA] triplets within the backbone, major, and minor grooves of DNA under different salt environments. Values are averaged over three independent replicas and calculated only within the bound-state intervals defined by the persistence-based criterion.

In Figure 8, the average number of [DNA–ion–PAA] triplets is resolved by region, distinguishing interactions involving the backbone, major groove, and minor groove. The results reveal distinct regional preferences that depend on the type of cation. In the NaCl system, triplet formation is minimal in all regions, confirming that $Na^+$ does not contribute significantly to direct coordination. In $MgCl_2$, bridging events occur primarily within the major and minor grooves, indicating that magnesium ions mediate more transient and spatially distributed contacts between the polymer and the DNA surface. In contrast, the $CaCl_2$ system exhibits a strong localization of bridging along the backbone, while groove-mediated triplets are absent. This distribution of PAA–DNA association toward the backbone is consistent with the larger number and higher persistence of $Ca^{2+}$ bridges observed previously (Table 3), showing that calcium ions promote more stable coordination directly at the phosphate groups. Together, our results clarify that although both divalent cations enhance DNA–PAA association compared to $Na^{+}$, only $Ca^{2+}$ induces a backbone-dominated binding mode that underlies the long-lived complexation seen in the simulations.

To interpret these regional differences, it is also useful to consider the known ion-binding preferences of DNA in the absence of PAA. Previous studies show that $Na^{+}$ accumulates most strongly in the minor groove[95,96], $Ca^{2+}$ binds preferentially along the phosphate backbone, and $Mg^{2+}$ is distributed between backbone and groove regions[96–98] but typically remains more hydrated and therefore less likely to form direct inner-sphere contacts. These intrinsic preferences explain the triplet patterns observed here. In the NaCl system, the minor-groove affinity of $Na^{+}$ does not translate into stable bridging because its monovalent charge and weak coordination prevent simultaneous contacts with DNA and PAA. In $CaCl_2$, the strong backbone affinity of $Ca^{2+}$, together with its more labile hydration shell, facilitates frequent direct coordination with phosphate groups, producing the dominant backbone-mediated bridges seen in our simulations. $Mg^{2+}$ occupies both backbone and groove regions in DNA-alone environments, yet its tightly bound hydration shell reduces its ability to form stable electrostatically mediated backbone bridges. Within the grooves, however, $Mg^{2+}$ can engage in hydrogen-bond–supported interactions with base-specific acceptor atoms, which provides a plausible mechanism for the groove-localized bridging observed in our $MgCl_2$ simulations.

In addition to these region-specific interaction patterns, the global geometry of the DNA chain was assessed through the asphericity, which measures how elongated or compact the overall DNA shape is. The calculation followed the S2, S3 and S4 equations listed in the Supporting Information. The asphericity distributions (Figure S7) show that NaCl and $CaCl_2$ maintain nearly identical coil shapes between DNA alone, unbound DNA–PAA segments, and bound intervals, indicating that these salts do not significantly distort the overall geometry of the DNA backbone. $MgCl_2$, however, produces a small but measurable shift toward lower asphericity during DNA–PAA binding, corresponding to a slightly less elongated overall shape. This behavior aligns with the groove-binding analysis, where $Mg^{2+}$ uniquely facilitates penetration of the major and minor grooves by PAA through ion-assisted charge compensation. Such groove interactions locally redistribute mass around the helical surface, enough to produce a modest reduction in elongation without altering the global size to the same extent as $Ca^{2+}$. This link between groove engagement and subtle shape adjustments in $Mg^{2+}$ systems is well-supported by previous studies reporting that $Mg^{2+}$ promotes closer DNA–ligand approach, partial groove insertion, and locally compacted configurations under multivalent ionic conditions[99,100].

## 4. CONCLUSION

This work investigates the salt dependent association between DNA and poly(acrylic acid) (PAA) using atomistic molecular dynamics simulations, with the aim of elucidating how two chemically dissimilar, like-charged polyelectrolytes can form stable complexes under different ionic conditions By accounting for ions with varying ion valency and ion solvation structure, the effects of specific counterion properties on the lifetime, overall structure and internal organization of DNA-PAA assemblies are investigated. A key aspect of our approach is the use of time resolved state classification, which allowed us to distinguish bound and unbound intervals and to characterize association as a dynamic, heterogeneous process rather than a single equilibrium state

The simulations reveal a clear hierarchy of complex stability across salts. CaCl2 produces long-lived association, whereas NaCl and MgCl2 display more frequent transitions and greater replica-to-replica variability. Ion–oxygen radial distribution functions show that $Ca^{2+}$ engages in inner-sphere coordination with DNA phosphates and PAA carboxylates. Consistent with this observation, direct bridge counting confirms that $Ca^{2+}$ acts as a molecular cross-linker, forming persistent connectors between the two anionic chains; $Mg^{2+}$ shows weaker, less frequent bridging, and $Na^{+}$ primarily screens without long-lived cross-links. The resulting polymer–polymer structuring follows the same ordering: the distribution of PAA oxygens around DNA phosphates is sharpest in $CaCl_2$, intermediate in $MgCl_2$, and diffuse in NaCl.

State-resolved structural metrics clarify how these microscopic interactions translate into conformational changes of the DNA. The RMSD of bound-DNA is largely unchanged in NaCl and $MgCl_2$ conditions relative to the reference case of DNA alone indicating that screening and transient contacts do not strongly perturb the DNA backbone. In $CaCl_2$, bound DNA shows slightly higher RMSD than both its unbound -state and the case of DNA alone, a result that is consistent with the constraints forming at the contact interface between DNA and PAA. The radius of gyration provides a complementary view:NaCl and $MgCl_2$ show limited state dependence and remain close to the reference DNA structure, whereas $CaCl_2$ displays a clear separation between bound and unbound states, with bound DNA adopting a more extended global size, while unbound DNA is comparatively compact. A consistent interpretation is that $Ca^{2+}$ -mediated cross-linking stiffens and aligns DNA segments within the complex, increasing the global size of DNA, despite the enhanced DNA-PAA association

To probe this reorganization beyond global size descriptors, the topology and orientation of the polyelectrolytes and complexes were examined through principal moments of inertia and relative axis alignment. The inertia analysis uncovers subtle yet systematic rearrangements in the mass distribution of both macromolecules, with binding in $Ca^{2+}$ solutions producing a more extended DNA shape and a compacted PAA conformation. This behavior reflects a cooperative response to strong cross-linking, leading to a tightly organized complex. The corresponding orientational distributions confirm that $Ca^{2+}$ drives the transition from flexible association to directional alignment, where the long axis of DNA and the short axis of PAA adopt a stable, low-angle configuration. The narrowing of orientational distributions under $Ca^{2+}$ indicates the emergence of well-defined topological order within the complex.

The spatial contact analysis provides the molecular basis for this organization. Persistent backbone association dominates in $Ca^{2+}$ environments, supported by dense cation-bridged triplets that lock the two anionic chains together. Groove binding, which appears under weaker ionic conditions, becomes secondary as $Ca^{2+}$ enforces a backbone-centered mode stabilized by multiple coordination links. These observations collectively define a structural hierarchy where $Ca^{2+}$-mediated bridges not only neutralize charges but also impose geometrical order, guiding both the compaction of the polymer and the expansion of the DNA framework within a single, coherent configuration.

Taken together, these results highlight several key points. First, multivalent ions are not interchangeable: differences in hydration structure, coordination behavior, and mobility lead to distinct mechanisms that directly influence DNA–PAA association. $Mg^{2+}$ has a small ionic radius, a rigid and tightly bound hydration shell, and low mobility, features that limit its ability to reorganize its local environment and approach both polymers simultaneously. $Ca^{2+}$, by contrast, has a more labile hydration structure, higher mobility, and a more flexible coordination environment, which together enable it to engage directly with phosphate and carboxylate oxygens.

This combination allows $Ca^{2+}$ to act as a cross-linker that stabilizes association and reorganizes short-range polymer contacts, while $Mg^{2+}$ provides moderate stabilization and $Na^{+}$ mainly screens without persistent bridging. Second, separating bound and unbound intervals is essential for systems that visit kinetically trapped ensembles, since single averages obscure the strong salt dependence revealed by state classification. Third, global size metrics must be interpreted together with contact patterns and ion coordination because localized backbone cross-linking can increase the apparent size of the DNA chain even when association is strong. Fourth, the consistent observation of association across all salts, despite the overall negative charges on both polymers, highlights the role of specific ion binding and bridging in overcoming simple electrostatic repulsion and shaping complex stability and structure. These results offer a molecular framework for understanding how ion valency and ionic size govern the self-assembly and stability of charged macromolecular systems, with relevance to gene delivery, biomaterial design, and the behavior of nucleic acids in crowded ionic environments. Future work that combines longer timescale simulations with experimental measurements can build on this foundation to guide the design of ion responsive materials and carriers that exploit controlled electrostatic association at the molecular scale.

## ASSOCIATED CONTENT

Twenty four base pair DNA duplex model used in simulations, Energy profiles and interaction energy components for NaCl, $MgCl_2$ and $CaCl_2$ systems, Bound fraction analysis across three replicas for each salt, Mean squared displacement of $Ca^{2+}$, $Mg^{2+}$ and $Na^{+}$ ions, Asphericity calculations and probability density distributions for DNA in bound, unbound and DNA only states, Time evolution of DNA–PAA contacts in backbone and grooves, Principal moments of inertia for DNA across salts and states, Principal moments of inertia for PAA across salts and states, state-resolved heavy-atom RMSD values of PAA. All figures S1 to S8 and Tables S1,S2 and S3.

## AUTHOR INFORMATION

### Corresponding Author

Terpsichori S. Alexiou, terpsichori.alexiou@univie.ac.at, Faculty of Physics, University of Vienna, Boltzmanngasse 5, A-1090, Vienna, Austria

### Author Contributions

The manuscript was written through contributions of all authors. All authors have given approval to the final version of the manuscript. ‡All authors contributed equally.

### Funding Sources

This project has received funding from the European Union's Horizon 2020 Research and Innovation Programme under the Marie Skłodowska-Curie Grant Agreement No. 101034267. T.A. acknowledges support from the Austrian Science Fund (FWF) [10.55776/RIC4065024].

### Conflict of Interest

The authors have no conflicts to disclose.

## REFERENCES

[1] V.S. Meka, M.K.G. Sing, M.R. Pichika, S.R. Nali, V.R.M. Kolapalli, and P. Kesharwani, "A comprehensive review on polyelectrolyte complexes," Drug Discovery Today **22**(11), 1697–1706 (2017).
[2] A. Dobrynin, and M. Rubinstein, "Theory of polyelectrolytes in solutions and at surfaces," Progress in Polymer Science **30**(11), 1049–1118 (2005).
[3] M. Ishihara, S. Kishimoto, S. Nakamura, Y. Sato, and H. Hattori, "Polyelectrolyte Complexes of Natural Polymers and Their Biomedical Applications," Polymers **11**(4), 672 (2019).

[4] A.D. Kulkarni, Y.H. Vanjari, K.H. Sancheti, H.M. Patel, V.S. Belgamwar, S.J. Surana, and C.V. Pardeshi, "Polyelectrolyte complexes: mechanisms, critical experimental aspects, and applications," Artificial Cells, Nanomedicine, and Biotechnology, (2016).
[5] Y. Shi, H. Peng, J. Yang, and J. Zhao, "Counterion Binding Dynamics of a Polyelectrolyte," Macromolecules **54**(10), 4926–4933 (2021).
[6] P. Jagtap, K. Patil, and P. Dhatrak, "Polyelectrolyte Complex for Drug Delivery in Biomedical Applications: A Review," IOP Conf. Ser.: Mater. Sci. Eng. **1183**(1), 012007 (2021).
[7] R. Cazorla-Luna, A. Martín-Illana, F. Notario-Pérez, R. Ruiz-Caro, and M.-D. Veiga, "Naturally Occurring Polyelectrolytes and Their Use for the Development of Complex-Based Mucoadhesive Drug Delivery Systems: An Overview," Polymers **13**(14), 2241 (2021).
[8] S. Lankalapalli, and V.R.M. Kolapalli, "Polyelectrolyte Complexes: A Review of their Applicability in Drug Delivery Technology," Indian J Pharm Sci **71**(5), 481–487 (2009).
[9] C.-J. Lee, H. Wu, Y. Hu, M. Young, H. Wang, D. Lynch, F. Xu, H. Cong, and G. Cheng, "Ionic Conductivity of Polyelectrolyte Hydrogels," ACS Appl. Mater. Interfaces **10**(6), 5845–5852 (2018).
[10] C.-G. Wang, N.E.B. Surat'man, J.J. Chang, Z.L. Ong, B. Li, X. Fan, X.J. Loh, and Z. Li, "Polyelectrolyte Hydrogels for Tissue Engineering and Regenerative Medicine," Chemistry – An Asian Journal **17**(18), e202200604 (2022).
[11] B. Bolto, and J. Gregory, "Organic polyelectrolytes in water treatment," Water Research **41**(11), 2301–2324 (2007).
[12] J.-P. Chapel, and J.-F. Berret, "Versatile electrostatic assembly of nanoparticles and polyelectrolytes: Coating, clustering and layer-by-layer processes," Current Opinion in Colloid & Interface Science **17**(2), 97–105 (2012).
[13] F.N. Ajjan, M. Ambrogi, G.A. Tiruye, D. Cordella, A.M. Fernandes, K. Grygiel, M. Isik, N. Patil, L. Porcarelli, G. Rocasalbas, G. Vendramientto, E. Zeglio, M. Antonietti, C. Detrembleur, O. Inganäs, C. Jérôme, R. Marcilla, D. Mecerreyes, M. Moreno, D. Taton, N. Solin, and J. Yuan, "Innovative polyelectrolytes/poly(ionic liquid)s for energy and the environment," Polymer International **66**(8), 1119–1128 (2017).
[14] G. Quek, B. Roehrich, Y. Su, L. Sepunaru, and G.C. Bazan, "Conjugated Polyelectrolytes: Underexplored Materials for Pseudocapacitive Energy Storage," Advanced Materials **34**(22), 2104206 (2022).
[15] R.S.T. Modderman, and L.W.J. Holleman, "Coacervation," Nature **129**(3261), 654–654 (1932).
[16] A.E. Neitzel, Y.N. Fang, B. Yu, A.M. Rumyantsev, J.J. de Pablo, and M.V. Tirrell, "Polyelectrolyte Complex Coacervation across a Broad Range of Charge Densities," Macromolecules **54**(14), 6878–6890 (2021).
[17] J.T.G. Overbeek, and M.J. Voorn, "Phase separation in polyelectrolyte solutions. Theory of complex coacervation," Journal of Cellular and Comparative Physiology **49**(S1), 7–26 (1957).
[18] C.B. Bucur, Z. Sui, and J.B. Schlenoff, "Ideal Mixing in Polyelectrolyte Complexes and Multilayers: Entropy Driven Assembly," J. Am. Chem. Soc. **128**(42), 13690–13691 (2006).
[19] R. Podgornik, "Polyelectrolyte-mediated bridging interactions," Journal of Polymer Science Part B: Polymer Physics **42**(19), 3539–3556 (2004).

[20] Q. Wang, and J.B. Schlenoff, "The Polyelectrolyte Complex/Coacervate Continuum," Macromolecules **47**(9), 3108–3116 (2014).
[21] F. Trilling, M.-K. Ausländer, and U. Scherf, "Ladder-Type Polymers and Ladder-Type Polyelectrolytes with On-Chain Dibenz[a,h]anthracene Chromophores," Macromolecules **52**(8), 3115–3122 (2019).
[22] A.A. Lazutin, A.N. Semenov, and V.V. Vasilevskaya, "Polyelectrolyte Complexes Consisting of Macromolecules With Varied Stiffness: Computer Simulation," Macromolecular Theory and Simulations **21**(5), 328–339 (2012).
[23] A. Naji, M. Kanduč, J. Forsman, and R. Podgornik, "Perspective: Coulomb fluids—Weak coupling, strong coupling, in between and beyond," J. Chem. Phys. **139**(15), 150901 (2013).
[24] I. Szilagyi, G. Trefalt, A. Tiraferri, P. Maroni, and M. Borkovec, "Polyelectrolyte adsorption, interparticle forces, and colloidal aggregation," (2014).
[25] A. Dobrynin, and M. Rubinstein, "Theory of polyelectrolytes in solutions and at surfaces," Progress in Polymer Science **30**(11), 1049–1118 (2005).
[26] V.A. Bloomfield, "Condensation of DNA by multivalent cations: Considerations on mechanism," Biopolymers **31**(13), 1471–1481 (1991).
[27] V. Lorman, R. Podgornik, and B. Žekš, "Positional, Reorientational, and Bond Orientational Order in DNA Mesophases," Phys. Rev. Lett. **87**(21), 218101 (2001).
[28] H.M. Harreis, A.A. Kornyshev, C.N. Likos, H. Löwen, and G. Sutmann, "Phase Behavior of Columnar DNA Assemblies," Phys. Rev. Lett. **89**(1), 018303 (2002).
[29] T.E. Angelini, R. Golestanian, R.H. Coridan, J.C. Butler, A. Beraud, M. Krisch, H. Sinn, K.S. Schweizer, and G.C.L. Wong, "Counterions between charged polymers exhibit liquid-like organization and dynamics," Proceedings of the National Academy of Sciences **103**(21), 7962–7967 (2006).
[30] C. Yigit, J. Heyda, M. Ballauff, and J. Dzubiella, "Like-charged protein-polyelectrolyte complexation driven by charge patches," J. Chem. Phys. **143**(6), 064905 (2015).
[31] S. Kim, J. Huang, Y. Lee, S. Dutta, H.Y. Yoo, Y.M. Jung, Y. Jho, H. Zeng, and D.S. Hwang, "Complexation and coacervation of like-charged polyelectrolytes inspired by mussels," Proceedings of the National Academy of Sciences **113**(7), E847–E853 (2016).
[32] S.L. Perry, Y. Li, D. Priftis, L. Leon, M. Tirrell, S.L. Perry, Y. Li, D. Priftis, L. Leon, and M. Tirrell, "The Effect of Salt on the Complex Coacervation of Vinyl Polyelectrolytes," Polymers **6**(6), 1756–1772 (2014).
[33] H. Yong, "Reentrant Condensation of Polyelectrolytes Induced by Diluted Multivalent Salts: The Role of Electrostatic Gluonic Effects," Biomacromolecules **25**(11), 7361–7376 (2024).
[34] T. Braide, S.M. Lalwani, C.I. Eneh, and J.L. Lutkenhaus, "Divalent cation effects in the glass transition of poly(diallyldimethylammonium)–poly(styrene sulfonate) polyelectrolyte complexes," Soft Matter **20**(48), 9631–9641 (2024).
[35] T.S. Alexiou, and C.N. Likos, "Ion-Specific Modulation of the Conformation and Compactness of DNA Oligo-Catenanes," J. Phys. Chem. B **130**(2), 868–880 (2026).
[36] W. Liu, A. Samanta, J. Deng, C.O. Akintayo, and A. Walther, "Mechanistic Insights into the Phase Separation Behavior and Pathway-Directed Information Exchange in all-DNA Droplets," Angewandte Chemie International Edition **61**(45), e202208951 (2022).

[37] A. Glisman, S. Mantha, D. Yu, E.P. Wasserman, S. Backer, and Z.-G. Wang, "Multivalent Ion-Mediated Polyelectrolyte Association and Structure," Macromolecules **57**(5), 1941–1949 (2024).
[38] N. Bali, S.J. Brennhaug, M. Bjørås, S. Bandyopadhyay, and A. Manaf, "Optimized synthesis of polyacrylic acid-coated magnetic nanoparticles for high-efficiency DNA isolation and size selection," RSC Adv. **13**(42), 29109–29120 (2023).
[39] H. Arkaban, M. Barani, M.R. Akbarizadeh, N. Pal Singh Chauhan, S. Jadoun, M. Dehghani Soltani, and P. Zarrintaj, "Polyacrylic Acid Nanoplatforms: Antimicrobial, Tissue Engineering, and Cancer Theranostic Applications," Polymers **14**(6), 1259 (2022).
[40] X. Hu, C.-J. Kim, S.K. Albert, and S.-J. Park, "DNA-Grafted Poly(acrylic acid) for One-Step DNA Functionalization of Iron Oxide Nanoparticles," Langmuir **34**(47), 14342–14346 (2018).
[41] Z. Min, B. Xu, W. Li, and A. Zhang, "Combination of DNA with polymers," Polym. Chem. **12**(13), 1898–1917 (2021).
[42] A.A. Anashkina, "Protein-DNA recognition mechanisms and specificity," Biophys Rev **15**(5), 1007–1014 (2023).
[43] E. Yekymov, D. Attia, Y. Levi-Kalisman, R. Bitton, R. Yerushalmi-Rozen, E. Yekymov, D. Attia, Y. Levi-Kalisman, R. Bitton, and R. Yerushalmi-Rozen, "Charge Regulation of Poly(acrylic acid) in Solutions of Non-Charged Polymer and Colloids," Polymers **15**(5), (2023).
[44] T. Swift, L. Swanson, M. Geoghegan, and S. Rimmer, "The pH-responsive behaviour of poly(acrylic acid) in aqueous solution is dependent on molar mass," Soft Matter **12**(9), 2542–2549 (2016).
[45] I. Ivani, P.D. Dans, A. Noy, A. Pérez, I. Faustino, A. Hospital, J. Walther, P. Andrio, R. Goñi, A. Balaceanu, G. Portella, F. Battistini, J.L. Gelpí, C. González, M. Vendruscolo, C.A. Laughton, S.A. Harris, D.A. Case, and M. Orozco, "Parmbsc1: a refined force field for DNA simulations," Nat Methods **13**(1), 55–58 (2016).
[46] C. Prior, L. Danilāne, and V. S. Oganesyan, "All-atom molecular dynamics simulations of spin labelled double and single-strand DNA for EPR studies," Physical Chemistry Chemical Physics **20**(19), 13461–13472 (2018).
[47] B. Knappeová, V. Mlýnský, M. Pykal, J. Šponer, P. Banáš, M. Otyepka, and M. Krepl, "Comprehensive Assessment of Force-Field Performance in Molecular Dynamics Simulations of DNA/RNA Hybrid Duplexes," J. Chem. Theory Comput. **20**(15), 6917–6929 (2024).
[48] P.D. Dans, I. Ivani, A. Hospital, G. Portella, C. González, and M. Orozco, "How accurate are accurate force-fields for B-DNA?," Nucleic Acids Res **45**(7), 4217–4230 (2017).
[49] D.G. Mintis, and V.G. Mavrantzas, "Effect of pH and Molecular Length on the Structure and Dynamics of Short Poly(acrylic acid) in Dilute Solution: Detailed Molecular Dynamics Study," J. Phys. Chem. B **123**(19), 4204–4219 (2019).
[50] J. Wang, R.M. Wolf, J.W. Caldwell, P.A. Kollman, and D.A. Case, "Development and testing of a general amber force field," Journal of Computational Chemistry **25**(9), 1157–1174 (2004).
[51] G.R. Quezada, N. Toro, J. Saavedra, P. Robles, I. Salazar, A. Navarra, R.I. Jeldres, G.R. Quezada, N. Toro, J. Saavedra, P. Robles, I. Salazar, A. Navarra, and R.I. Jeldres, "Molecular Dynamics Study of the Conformation, Ion Adsorption, Diffusion, and Water Structure of Soluble Polymers in Saline Solutions," Polymers **13**(20), (2021).

[52] W.L. Jorgensen, J. Chandrasekhar, J.D. Madura, R.W. Impey, and M.L. Klein, “Comparison of simple potential functions for simulating liquid water,” The Journal of Chemical Physics **79**(2), 926–935 (1983).
[53] I.S. Joung, and T.E.I. Cheatham, “Determination of Alkali and Halide Monovalent Ion Parameters for Use in Explicitly Solvated Biomolecular Simulations,” J. Phys. Chem. B **112**(30), 9020–9041 (2008).
[54] I.S. Joung, and T.E.I. Cheatham, “Molecular Dynamics Simulations of the Dynamic and Energetic Properties of Alkali and Halide Ions Using Water-Model-Specific Ion Parameters,” J. Phys. Chem. B **113**(40), 13279–13290 (2009).
[55] T. Darden, D. York, and L. Pedersen, “Particle mesh Ewald: An $N \cdot \log(N)$ method for Ewald sums in large systems,” The Journal of Chemical Physics **98**(12), 10089–10092 (1993).
[56] X. Xu, S. Angioletti-Uberti, Y. Lu, J. Dzubiella, and M. Ballauff, “Interaction of Proteins with Polyelectrolytes: Comparison of Theory to Experiment,” Langmuir **35**(16), 5373–5391 (2019).
[57] J. Bukala, P. Yavvari, J. Walkowiak, M. Ballauff, and M. Weinhart, “Interaction of Linear Polyelectrolytes with Proteins: Role of Specific Charge–Charge Interaction and Ionic Strength,” Biomolecules **11**(9), 1377 (2021).
[58] A. Clarke, K. Groschner, and T. Stockner, “Exploring TRPC3 Interaction with Cholesterol through Coarse-Grained Molecular Dynamics Simulations,” Biomolecules **12**(7), 890 (2022).
[59] Y. Liu, Y. Liu, M.B. Chan-Park, and Y. Mu, “Binding Modes of Teixobactin to Lipid II: Molecular Dynamics Study,” Sci Rep **7**(1), 17197 (2017).
[60] Í.P. Caruso, K. Sanches, A.T. Da Poian, A.S. Pinheiro, and F.C.L. Almeida, “Dynamics of the SARS-CoV-2 nucleoprotein N-terminal domain triggers RNA duplex destabilization,” Biophysical Journal **120**(14), 2814–2827 (2021).
[61] A. Bartocci, A. Grazzi, N. Awad, P.-J. Corringer, P.C.T. Souza, and M. Cecchini, “A millisecond coarse-grained simulation approach to decipher allosteric cannabinoid binding at the glycine receptor α1,” Nat Commun **15**(1), 9040 (2024).
[62] E. Schulze-Niemand, M. Naumann, and M. Stein, “Substrate-assisted activation and selectivity of the bacterial RavD effector deubiquitinylase,” Proteins: Structure, Function, and Bioinformatics **90**(4), 947–958 (2022).
[63] G.P. Costa, S.R. Stoyanov, Q. Liu, and P. Choi, “Molecular Dynamics Simulation of the Adsorption Interactions of Selected Polar and Nonpolar Polymers on Kaolinite Basal Surfaces in the Presence of Cyclohexane,” The Journal of Physical Chemistry C, (2025).
[64] A. Luzar, and D. Chandler, “Hydrogen-bond kinetics in liquid water,” Nature **379**(6560), 55–57 (1996).
[65] I.K. McDonald, and J.M. Thornton, “Satisfying hydrogen bonding potential in proteins,” J Mol Biol **238**(5), 777–793 (1994).
[66] D.F. Stickle, L.G. Presta, K.A. Dill, and G.D. Rose, “Hydrogen bonding in globular proteins,” J Mol Biol **226**(4), 1143–1159 (1992).
[67] A. Villa, J. Wöhnert, and G. Stock, “Molecular dynamics simulation study of the binding of purine bases to the aptamer domain of the guanine sensing riboswitch,” Nucleic Acids Res **37**(14), 4774–4786 (2009).

[68] F. Guerra, M. Siemers, C. Mielack, and A.-N. Bondar, "Dynamics of Long-Distance Hydrogen-Bond Networks in Photosystem II," The Journal of Physical Chemistry B, (2018).
[69] K.P. Tan, K. Singh, A. Hazra, and M.S. Madhusudhan, "Peptide bond planarity constrains hydrogen bond geometry and influences secondary structure conformations," Curr Res Struct Biol **3**, 1–8 (2020).
[70] J. Liu, X. He, J.Z. H. Zhang, and L.-W. Qi, "Hydrogen-bond structure dynamics in bulk water: insights from ab initio simulations with coupled cluster theory," Chemical Science **9**(8), 2065–2073 (2018).
[71] F.W. Starr, "Fast and Slow Dynamics of Hydrogen Bonds in Liquid Water," Phys. Rev. Lett. **82**(11), 2294–2297 (1999).
[72] S. Sen, D. Andreatta, S.Y. Ponomarev, D.L. Beveridge, and M.A. Berg, "Dynamics of water and ions near DNA: comparison of simulation to time-resolved stokes-shift experiments," J Am Chem Soc **131**(5), 1724–1735 (2009).
[73] H. Shweta, and S. Sen, "Dynamics of water and ions around DNA: What is so special about them?," J Biosci **43**(3), 499–518 (2018).
[74] S.Y. Ponomarev, K.M. Thayer, and D.L. Beveridge, "Ion motions in molecular dynamics simulations on DNA," Proceedings of the National Academy of Sciences **101**(41), 14771–14775 (2004).
[75] A.Yu. Antipina, and A.A. Gurtovenko, "Molecular Mechanism of Calcium-Induced Adsorption of DNA on Zwitterionic Phospholipid Membranes," J. Phys. Chem. B **119**(22), 6638–6645 (2015).
[76] A. Pérez, F.J. Luque, and M. Orozco, "Dynamics of B-DNA on the Microsecond Time Scale," J. Am. Chem. Soc. **129**(47), 14739–14745 (2007).
[77] T.E. Cheatham, "Simulation and modeling of nucleic acid structure, dynamics and interactions," Current Opinion in Structural Biology **14**(3), 360–367 (2004).
[78] M. Yang, J. Shi, and J.B. Schlenoff, "Control of Dynamics in Polyelectrolyte Complexes by Temperature and Salt," (2019).
[79] A. Chremos, and J.F. Douglas, "Polyelectrolyte association and solvation," J Chem Phys **149**(16), 163305 (2018).
[80] I. Bos, and J. Sprakel, "Langevin Dynamics Simulations of the Exchange of Complex Coacervate Core Micelles: The Role of Nonelectrostatic Attraction and Polyelectrolyte Length," Macromolecules **52**(22), 8923–8931 (2019).
[81] J. Li, and K. Zhao, "Effect of Side-Chain on Conformation of Poly(acrylic acid) and Its Dielectric Behaviors in Aqueous Solution: Hydrophobic and Hydrogen-Bonding Interactions and Mechanism of Relaxations," J. Phys. Chem. B **117**(39), 11843–11852 (2013).
[82] S.M. Lalwani, P. Batys, M. Sammalkorpi, and J.L. Lutkenhaus, "Relaxation Times of Solid-like Polyelectrolyte Complexes of Varying pH and Water Content," Macromolecules **54**(17), 7765–7776 (2021).
[83] O. Allnér, L. Nilsson, and A. Villa, "Magnesium Ion–Water Coordination and Exchange in Biomolecular Simulations," J. Chem. Theory Comput. **8**(4), 1493–1502 (2012).
[84] H. Wu, J.M. Ting, O. Werba, S. Meng, and M.V. Tirrell, "Non-equilibrium phenomena and kinetic pathways in self-assembled polyelectrolyte complexes," J Chem Phys **149**(16), 163330 (2018).

[85] D. Zhuang, M. Riera, R. Zhou, A. Deary, and F. Paesani, "Hydration Structure of Na+ and K+ Ions in Solution Predicted by Data-Driven Many-Body Potentials," (n.d.).
[86] A.A.A. Delgado, D. Sethio, E. Kraka, A.A.A. Delgado, D. Sethio, and E. Kraka, "Assessing the Intrinsic Strengths of Ion–Solvent and Solvent–Solvent Interactions for Hydrated Mg2+ Clusters," Inorganics **9**(5), (2021).
[87] R. Shi, Z. Zhao, X. Huang, P. Wang, Y. Su, L. Sai, X. Liang, H. Han, and J. Zhao, "Ground-State Structures of Hydrated Calcium Ion Clusters From Comprehensive Genetic Algorithm Search," Front. Chem. **9**, (2021).
[88] A. Srivastava, R. Timsina, S. Heo, S.W. Dewage, S. Kirmizialtin, and X. Qiu, "Structure-guided DNA–DNA attraction mediated by divalent cations," Nucleic Acids Res **48**(13), 7018–7026 (2020).
[89] J. Yoo, and A. Aksimentiev, "Competitive Binding of Cations to Duplex DNA Revealed through Molecular Dynamics Simulations," J. Phys. Chem. B **116**(43), 12946–12954 (2012).
[90] P. Li, and K.M. Merz, "Metal Ion Modeling Using Classical Mechanics," Chem Rev **117**(3), 1564–1686 (2017).
[91] I.-C. Yeh, and G. Hummer, "System-Size Dependence of Diffusion Coefficients and Viscosities from Molecular Dynamics Simulations with Periodic Boundary Conditions," J. Phys. Chem. B **108**(40), 15873–15879 (2004).
[92] T. Ikeda, M. Boero, and K. Terakura, "Hydration properties of magnesium and calcium ions from constrained first principles molecular dynamics," The Journal of Chemical Physics **127**(7), 074503 (2007).
[93] S. Guilbaud, "Dependence of DNA Persistence Length on Ionic Strength and Ion Type," Phys. Rev. Lett. **122**(2), (2019).
[94] R.R. Netz, and H. Orland, "Variational charge renormalization in charged systems," Eur. Phys. J. E **11**(3), 301–311 (2003).
[95] F. Cesare Marincola, V.P. Denisov, and B. Halle, "Competitive Na+ and Rb+ Binding in the Minor Groove of DNA," J. Am. Chem. Soc. **126**(21), 6739–6750 (2004).
[96] N.V. Hud, and M. Polak, "DNA–cation interactions: the major and minor grooves are flexible ionophores," Current Opinion in Structural Biology **11**(3), 293–301 (2001).
[97] M. Egli, "DNA-Cation Interactions: Quo Vadis?," Chemistry & Biology **9**(3), 277–286 (2002).
[98] B.P. Fingerhut, J. Schauss, A. Kundu, and T. Elsaesser, "Contact pairs of RNA with magnesium ions-electrostatics beyond the Poisson-Boltzmann equation," Biophysical Journal **120**(23), 5322–5332 (2021).
[99] S. Bae, I. Oh, J. Yoo, and J.S. Kim, "Effect of DNA Flexibility on Complex Formation of a Cationic Nanoparticle with Double-Stranded DNA," ACS Omega **6**(29), 18728–18736 (2021).
[100] M. Guéroult, O. Boittin, O. Mauffret, C. Etchebest, and B. Hartmann, "Mg2+ in the Major Groove Modulates B-DNA Structure and Dynamics," PLOS ONE **7**(7), e41704 (2012).

Supplementary Material For:

# Ion-Modulated Polyelectrolyte Complexation of DNA and Polyacrylic Acid from Molecular Dynamics Simulations

*Sisem Ektirici[1], Vagelis Harmandaris[1,2,3], Christos N. Likos[4], Terpsichori S. Alexiou[4]*

1. Computation-Based Science and Technology Research Center, The Cyprus Institute, Nicosia 2121, Cyprus
2. Department of Mathematics and Applied Mathematics, University of Crete, GR-71409 Heraklion, Greece
3. Institute of Applied and Computational Mathematics, Foundation for Research and Technology Hellas, IACM/FORTH, GR-71110 Heraklion, Greece
4. Faculty of Physics, University of Vienna, Boltzmanngasse 5, A-1090, Vienna, Austria

**DNA Model**

In order to see the precise DNA construct used throughout the atomistic simulation study in a transparent way, the full twenty four base pair duplex is shown so the reader can clearly identify the nucleotide sequence, the base pair pattern, and the exact structural motif that served as the starting point of all simulations. Figure S1 provides an explicit reference for the composition of the DNA model and helps contextualize the subsequent analyses carried out on the system.

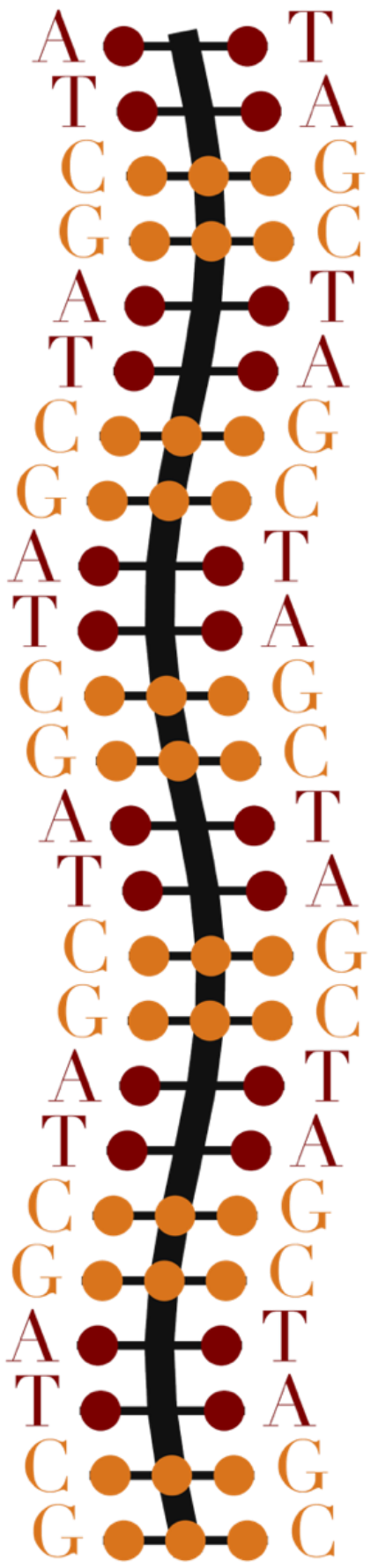

**Figure S1.** Representation of the 24-bp DNA duplex used in this study (5′–ATCGATCGATCGATCGATCGATCG–3′), illustrating Watson–Crick base pairing. Hydrogen bonds are shown as two dots for A–T base pairs and three dots for C–G base pairs.

**Energy Profiles of the Systems and Bound Fraction Analysis**

We calculated the energy profiles and the bound fraction analysis for the DNA–PAA complexes under different ionic conditions. Figure S2 shows the time evolution of the Short Range-Lennard-Jones (SR-$E_{LJ}$), Short Range-Coulomb (SR-$E_{Coul}$), and total interaction energies ($E_{Total}$) for NaCl, $MgCl_2$, and $CaCl_2$ systems across three independent replicas. Complementing this, Figures S3, S4, and S5 show the bound fraction analysis, illustrating the fraction of polymer chains closely associated with DNA over the simulation time.

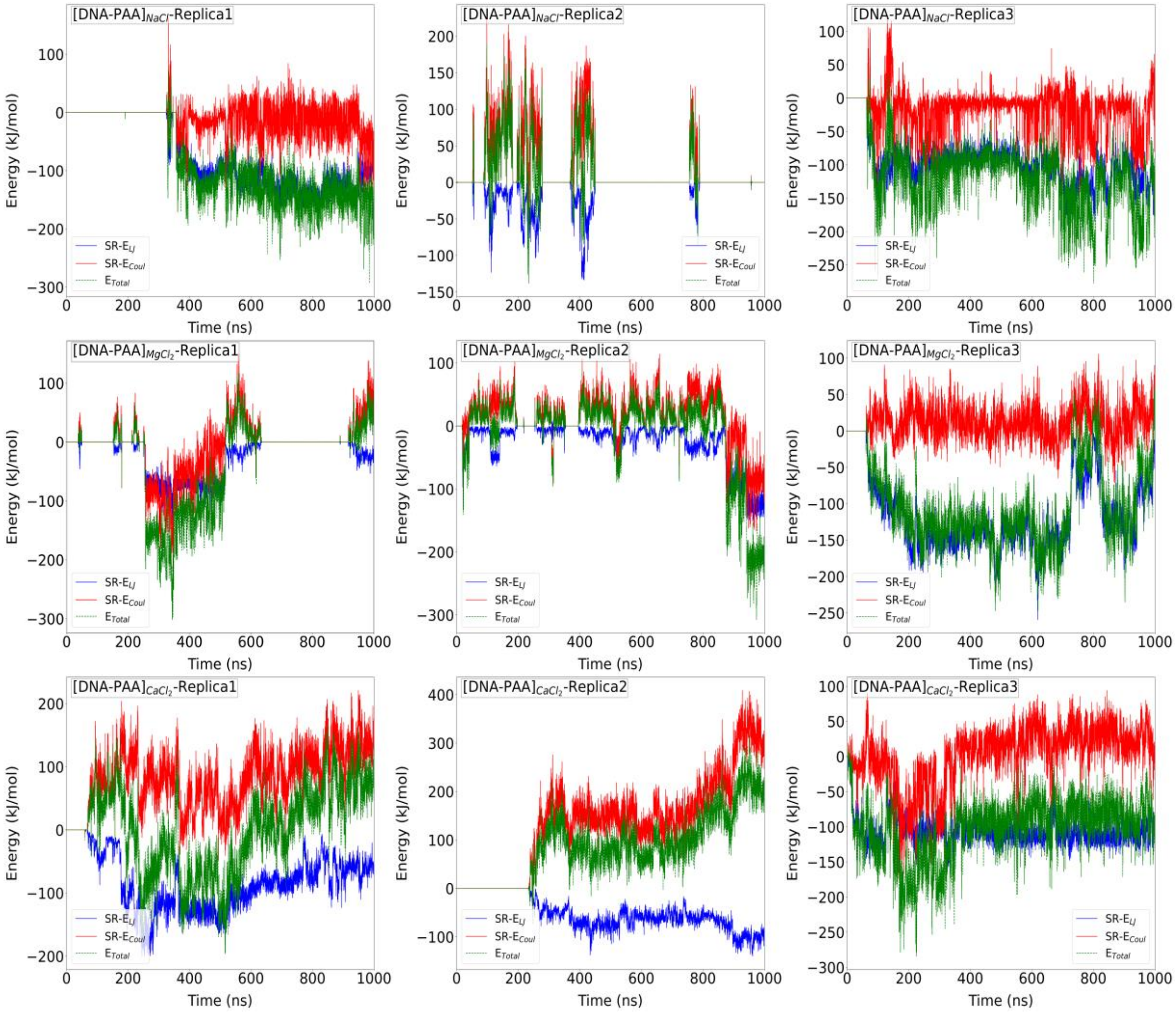

**Figure S2.** Time evolution of SR-Lennard-Jones (SR-$E_{LJ}$), SR-Coulomb (SR-$E_{Coul}$), and total interaction energies ($E_{Total}$) for DNA–PAA complexes in NaCl, $MgCl_2$, and $CaCl_2$ solutions across three replicas.

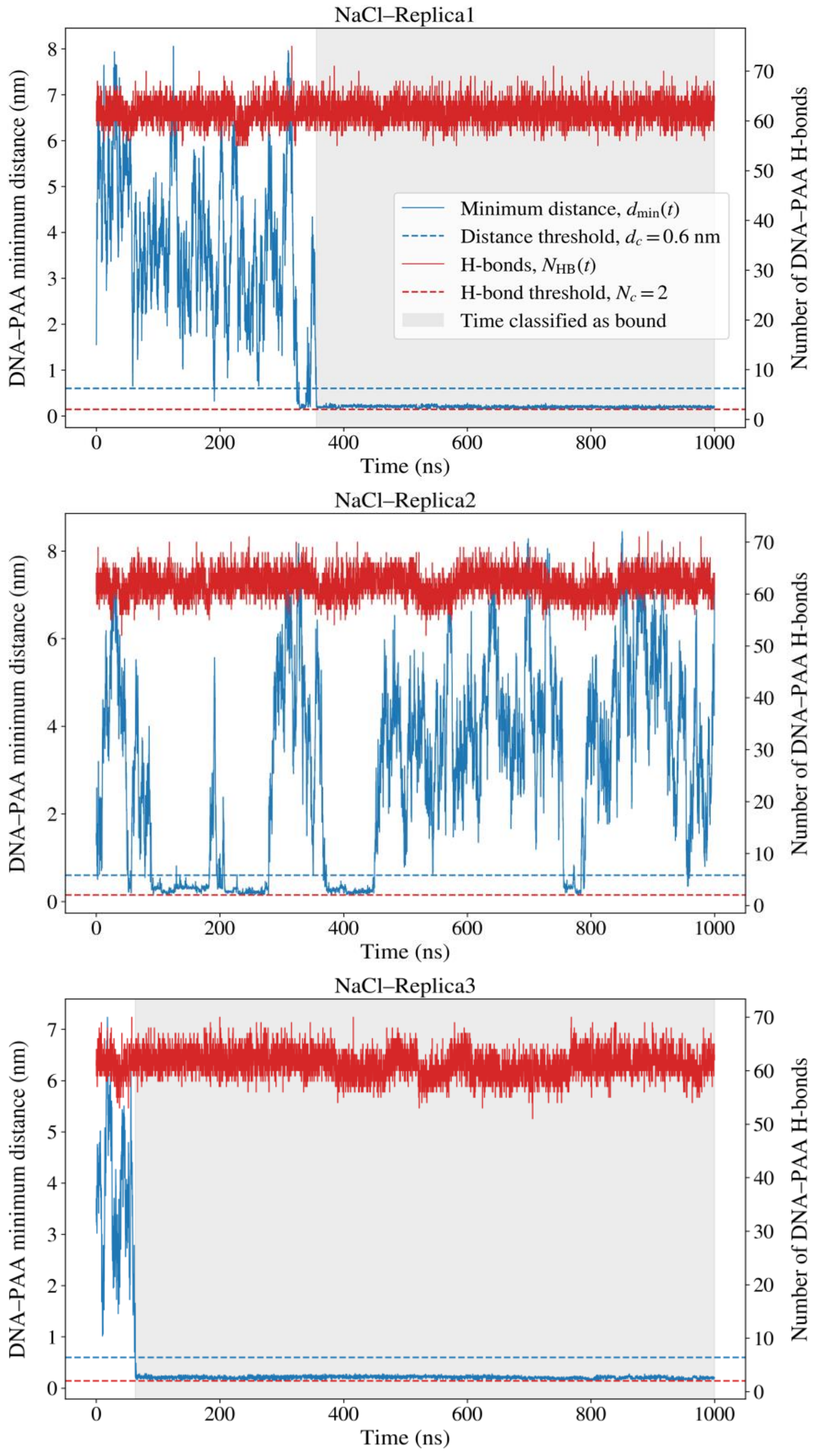

**Figure S3.** Minimum DNA–PAA distance and number of hydrogen bonds over time for three independent NaCl simulations, with persistent binding regions highlighted.

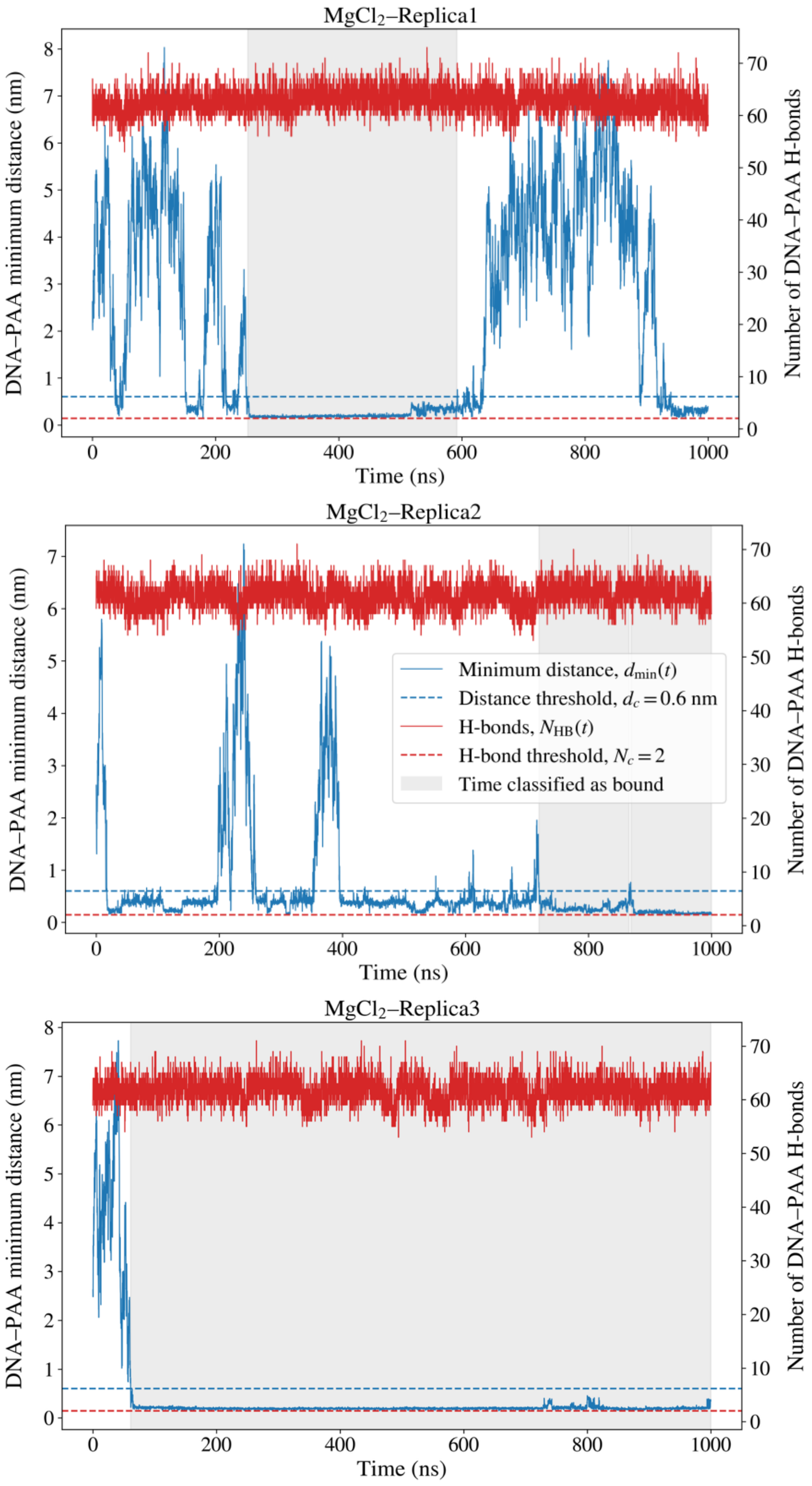

**Figure S4.** Minimum DNA–PAA distance and number of hydrogen bonds over time for three independent MgCl2 simulations, with persistent binding regions highlighted.

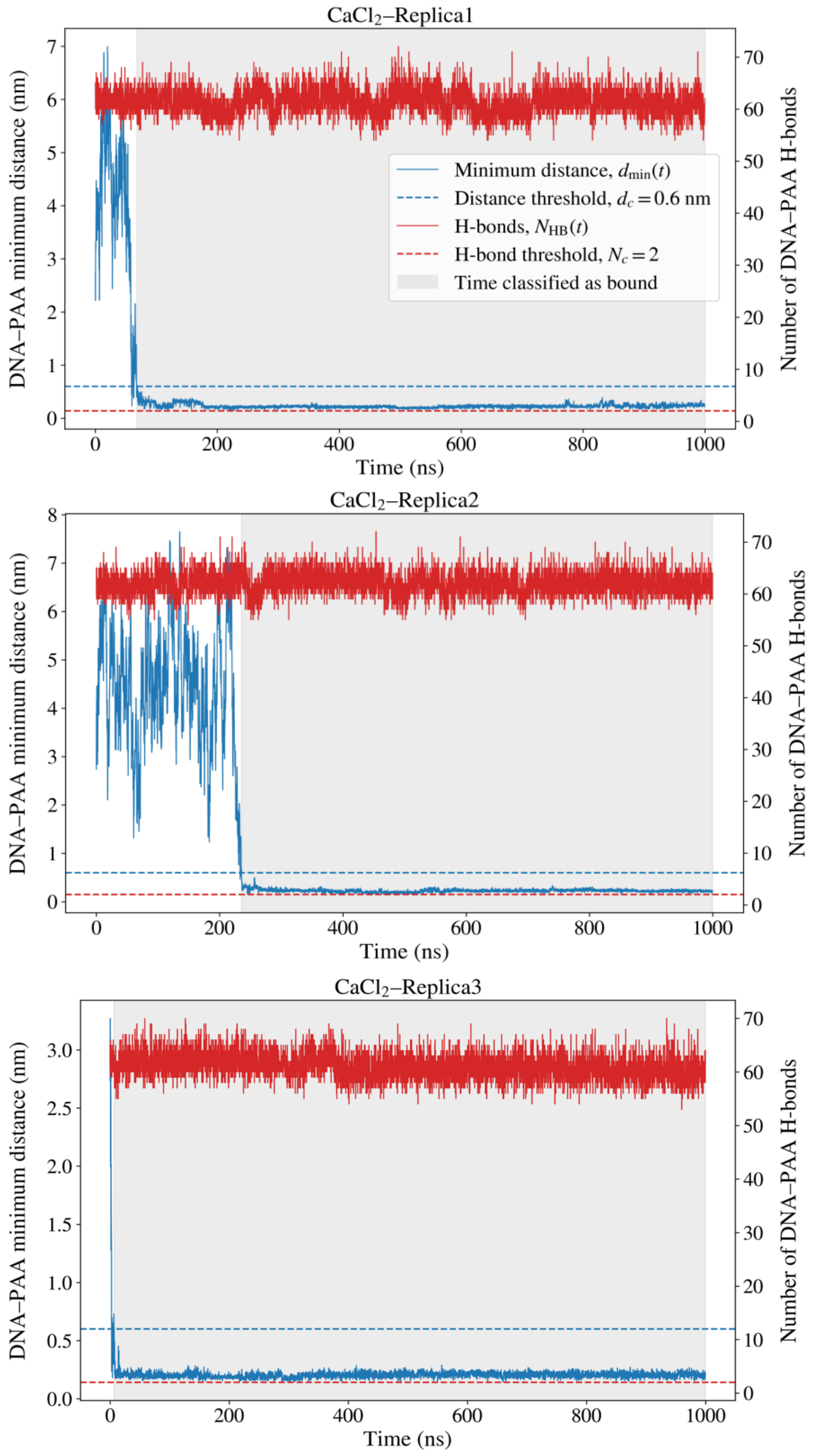

**Figure S5.** Minimum DNA–PAA distance and number of hydrogen bonds over time for three independent $CaCl_2$ simulations, with persistent binding regions highlighted.

**Mean Square Displacement (MSD) of Ions**

We evaluated the mobility of the ions in the simulation environment in order to understand how different cation types influence the dynamical properties of the system. The mean squared displacement provided a quantitative measure of ion diffusion and allowed us to compare the relative mobility of $Ca^{2+}$, $Mg^{2+}$ and $Na^{+}$ under identical conditions. MSD was computed using Equation S1:

$$MSD(t) = \frac{1}{N}\sum_{i} |r_i(t + t_0) - r_i(t_0)|^2 \qquad \text{S1}$$

where $r_i(t)$ is the position of ion i at time t and averaging was performed over all ions and all time origins. MSD values for ions are shown in Figure S6.

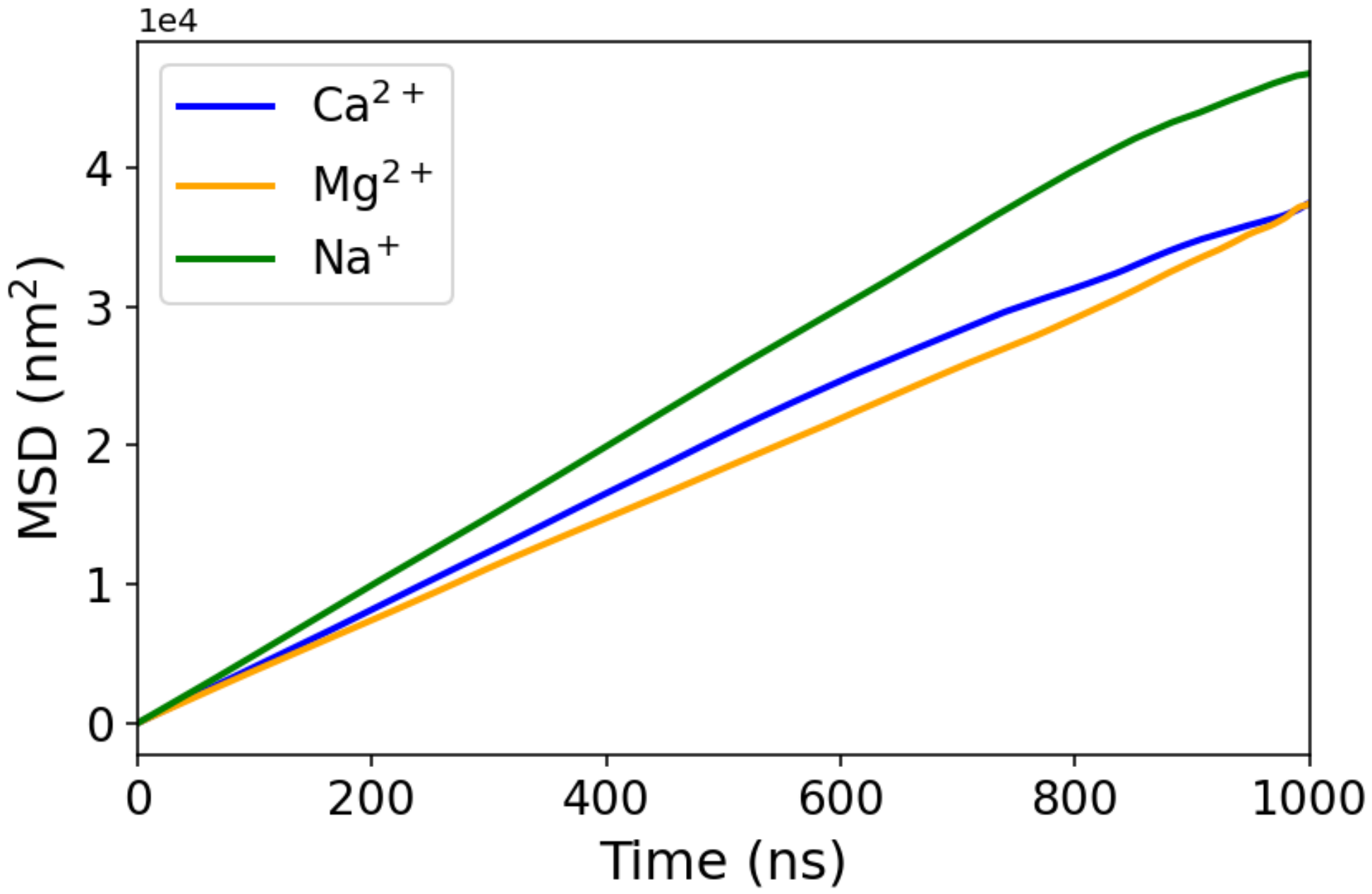

**Figure S6.** Mean squared displacement (MSD) of $Ca^{2+}$, $Mg^{2+}$ and $Na^{+}$ ions averaged over three simulation replicas.

**Asphericity**

To quantify shape differences of DNA under different salt conditions, we computed the normalized asphericity (A), from the eigenvalues of the gyration tensor. For each trajectory frame, the gyration tensor was constructed from the atomic coordinates and diagonalized to obtain the three principal eigenvalues λx, λy, and λz, ordered as λx ≤ λy ≤ λz. Following the formulation of Rudnick and Gaspari[1], the quantities $T_1$, $T_2$ and A were computed as:

$$T_1 = \lambda x \, \lambda y + \lambda y \, \lambda z + \lambda z \, \lambda x \qquad \text{S2}$$

$$T_2 = \lambda x + \lambda y + \lambda z \qquad \text{S3}$$

$$A = 1 - 3\, T_1 / T_2^2 \qquad \text{S4}$$

Asphericity values were computed separately for the bound and unbound segments of each replica by using Equations S2, S3 and S4. Replica-averaged values were obtained by calculating the mean asphericity for each replica and then averaging these values across the replicas that contained the corresponding state, with uncertainties reported as the standard error. DNA-only systems, which do not form polymer contacts, were treated as entirely unbound and analyzed in the same manner. For each salt condition, the distributions of asphericity values were characterized through probability density plots as shown in Figure S7. Kernel density estimates were computed for the bound, unbound, and DNA-only ensembles, normalized to unit height to enable direct comparison of distribution shapes. Figure S7 illustrate how DNA asphericity differs between polymer-bound and unbound states and how these states compare to DNA alone.

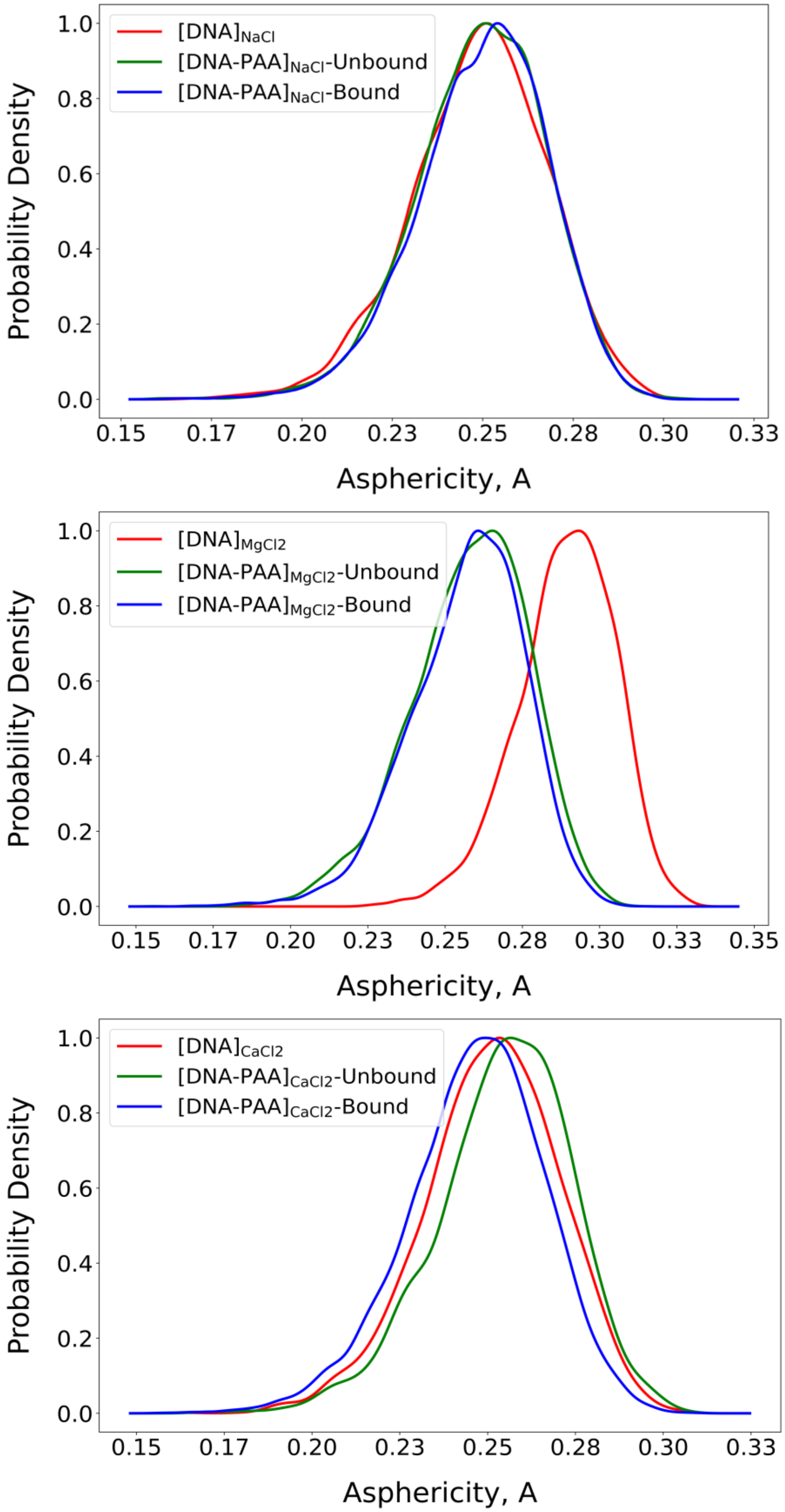

**Figure S7.** Comparison of DNA asphericity probability densities in NaCl, $MgCl_2$, and $CaCl_2$ systems showing free DNA and DNA–PAA complexes in unbound and bound states.

## Structural and Contact Analysis of DNA-PAA Complexes

We calculated the principal moments of inertia ($I_1$, $I_2$, $I_3$) for DNA and PAA in bound and unbound states to characterize changes in molecular shape and mass distribution under different ionic environments. The results, summarized in Tables S1 and S2 show a quantitative basis for comparing topological rearrangements across salts and replicas. Figure S8 reports the time evolution of DNA–PAA contacts along the backbone and within the grooves during bound intervals.

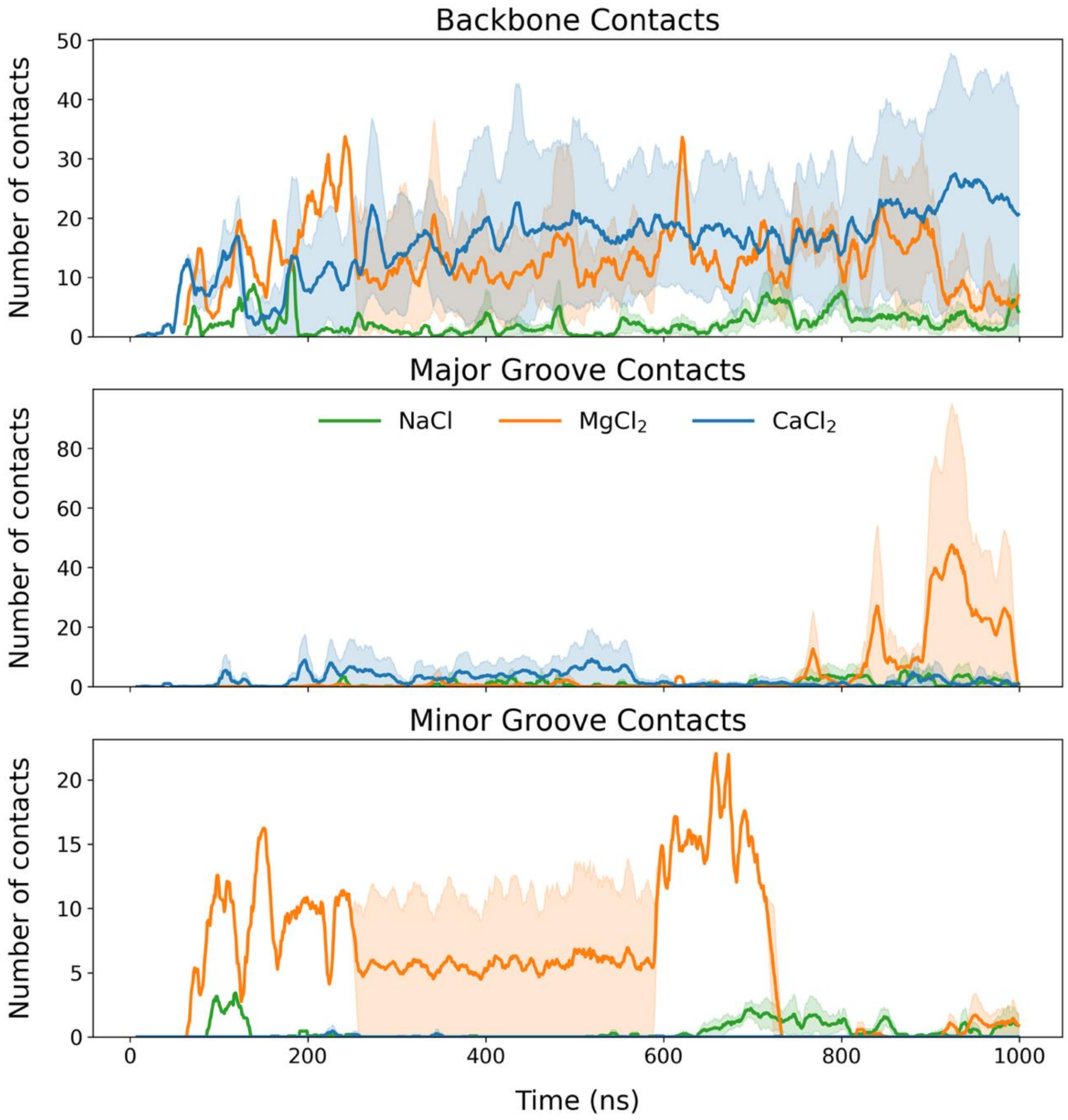

**Figure S8.** Time evolution of DNA–PAA contacts in the backbone and grooves under different salt environments during bound intervals.

**Table S1.** Principal moments of inertia I1, I2, I3 for DNA in bound and unbound states across different salt conditions and replicas, reported in amu·nm².

| **Salt and Replica** | **State** | **$I_1$ amu·nm²** | **$I_2$ amu·nm²** | **$I_3$ amu·nm²** |
|---|---|---|---|---|
| $CaCl_2$-1 | Bound | 12085.5 | 162136.8 | 165712.3 |
| $CaCl_2$-1 | Unbound | 9433.0 | 88430.9 | 89704.5 |
| $CaCl_2$-2 | Bound | 12878.1 | 155585.6 | 159976.2 |
| $CaCl_2$-2 | Unbound | 11195.1 | 149800.1 | 152831.0 |
| $CaCl_2$-3 | Bound | 10856.3 | 121732.2 | 124328.1 |
| $CaCl_2$-3 | Unbound | 9116.9 | 91487.5 | 92546.2 |
| **$CaCl_2$-Mean** | Bound | 11940.0±1018.7 | 146484.9±21685.2 | 150005.5±22421.4 |
| **$CaCl_2$-Mean** | Unbound | 9915.0 ± 1119.7 | 109906.2±34582.9 | 111693.9±35654.0 |
| $MgCl_2$-1 | Bound | 9545.1 | 90876.6 | 92333.8 |
| $MgCl_2$-1 | Unbound | 10520.6 | 120112.4 | 122528.8 |
| $MgCl_2$-2 | Bound | 11473.1 | 133176.3 | 136590.2 |
| $MgCl_2$-2 | Unbound | 12910.3 | 157018.9 | 161937.5 |
| $MgCl_2$-3 | Bound | 11696.0 | 139287.8 | 142966.8 |
| $MgCl_2$-3 | Unbound | 11027.8 | 154581.2 | 157662.9 |
| **$MgCl_2$-Mean** | Bound | 10904.5±1182.8 | 121108.3±26363.7 | 123958.2±27577.2 |
| **$MgCl_2$-Mean** | Unbound | 11488.0± 1258.0 | 143969.6±20683.1 | 147442.8±21663.9 |
| NaCl-1 | Bound | 11629.1 | 137512.2 | 140783.7 |
| NaCl-1 | Unbound | 13180.6 | 182325.4 | 187073.1 |
| NaCl-2 | Bound | - | - | - |
| NaCl-2 | Unbound | 11802.6 | 160281.1 | 163687.4 |
| NaCl-3 | Bound | 12378.6 | 174120.5 | 178139.6 |
| NaCl-3 | Unbound | 9555.0 | 122229.0 | 123527.7 |
| **NaCl-Mean** | Bound | 12003.9 ± 529.9 | 155816.3±25886.0 | 159461.6±26414.5 |
| **NaCl-Mean** | Unbound | 11512.7±1830.1 | 154945.2±30401.4 | 158096.1±32139.5 |

**Table S2.** Principal moments of inertia (I1, I2, I3) for PAA in bound and unbound states across different salt conditions and replicas, reported in amu·nm².

| **Salt and Replica** | **State** | **$I_1$ amu·nm²** | **$I_2$ amu·nm²** | **$I_3$ amu·nm²** |
|---|---|---|---|---|
| $CaCl_2$-1 | Bound | 1235.3 | 2345.1 | 2916.5 |
| $CaCl_2$-1 | Unbound | 1261.7 | 3071.6 | 3775.8 |
| $CaCl_2$-2 | Bound | 985.6 | 2760.9 | 3376.2 |
| $CaCl_2$-2 | Unbound | 996.0 | 3484.6 | 3976.9 |
| $CaCl_2$-3 | Bound | 1043.3 | 2077.5 | 2419.8 |
| $CaCl_2$-3 | Unbound | 1318.8 | 4555.9 | 5261.8 |
| **$CaCl_2$-Mean** | Bound | 1088.1±130.7 | 2394.5±344.3 | 2904.2± 478.3 |
| **$CaCl_2$-Mean** | Unbound | 1192.1±172.28 | 3704.0±766.1 | 4338.2±806.1 |
| $MgCl_2$-1 | Bound | 877.5 | 3031.4 | 3400.1 |
| $MgCl_2$-1 | Unbound | 920.5 | 2629.3 | 2938.7 |
| $MgCl_2$-2 | Bound | 960.2 | 1536.4 | 1927.4 |
| $MgCl_2$-2 | Unbound | 890.1 | 1902.7 | 2155.9 |
| $MgCl_2$-3 | Bound | 1232.6 | 3410.0 | 3919.6 |
| $MgCl_2$-3 | Unbound | 967.3 | 2357.2 | 2660.2 |
| **$MgCl_2$-Mean** | Bound | 1023.4±185.7 | 2659.2± 990.5 | 3082.3±1033.2 |
| **$MgCl_2$-Mean** | Unbound | 925.97 ± 38.81 | 2297.2± 367.2 | 2585.9 ±396.9 |
| NaCl-1 | Bound | 913.8 | 2303.9 | 2620.9 |
| NaCl-1 | Unbound | 1091.7 | 2605.1 | 2951.3 |
| NaCl-2 | Bound | - | - | - |
| NaCl-2 | Unbound | 1021.0 | 1658.0 | 2059.2 |
| NaCl-3 | Bound | 1243.5 | 1992.4 | 2482.0 |
| NaCl-3 | Unbound | 1137.5 | 1763.5 | 2096.0 |
| **NaCl-Mean** | Bound | 1078.7± 233.1 | 2148.2± 220.2 | 2551.4± 98.2 |
| **NaCl-Mean** | Unbound | 1083.4± 58.6 | 2008.9± 519.0 | 2368.9±504.7 |

**RMSD of Poly(acrylicacid)**

In this section, we assess the structural response of the PAA to DNA association by comparing Root Mean Square Deviation (RMSD) in the bound and unbound states. State-resolved heavy-RMSD values for PAA are summarized in Tables S3.

**Table S3.** State-resolved heavy-atom RMSD (nm) by salt for DNA alone and DNA–PAA systems.

| Salt | $RMSD_{bound}$ (nm) | $RMSD_{unbound}$ (nm) | $RMSD_{overall}$ (nm) | $RMSD_{[PAA]alone}$ (nm) |
|---|---|---|---|---|
| NaCl | 1.2 ± 0.06 | 1.2 ± 0.06 | 1.2 ± 0.07 | 1.0 ± 0.01 |
| MgCl2 | 1.07 ± 0.08 | 1.0 ± 0.14 | 1.03 ± 0.08 | 1.0 ± 0.02 |
| CaCl2 | 1.01 ± 0.06 | 0.8 ± 0.09 | 1.0 ± 0.06 | 1.04 ± 0.002 |

## REFERENCES

[1] J. Rudnick, and G. Gaspari, "The Shapes of Random Walks," Science **237**(4813), 384–389 (1987).